\documentclass[twocolumn]{aastex631}

\usepackage{threeparttable}
\usepackage{parskip}
\usepackage{float}
\usepackage{hyperref}
\usepackage{graphicx}
\usepackage{comment}

\usepackage{placeins}
\usepackage{booktabs}

\usepackage{amsthm}
\usepackage{amsmath}

\usepackage{soul}
\shorttitle{X-ray Variability of Mrk 421 with XPoSat and NICER}
\shortauthors{Akbar et al.}

\begin{document}
\title{X-ray Variability of Mrk 421 in 2025 January: A Ten-Day XPoSat XSPECT
Campaign with Contemporaneous NICER Coverage}

\correspondingauthor{Zahir Shah}

\author{Sikandar Akbar}
\affiliation{Department of Physics, University of Kashmir, Srinagar 190006, India.}

\author{Zahir Shah}
\affiliation{Manipal Centre for Natural Sciences, Manipal Academy of Higher Education, Manipal, India}
\email{zahir.shah@manipal.edu}

\author{Zahoor Malik}
\affiliation{Department of Physics University of the Free State Bloemfontein 9301 South Africa.}

\author{Faheem Aalijah }
\affiliation{Department of Physics, University of Kashmir, Srinagar 190006, India.}

\author{P Sreekumar}
\affiliation{Manipal Centre for Natural Sciences, Manipal Academy of Higher Education, Manipal, India}

\author{Vivek Kumar}
\affiliation{U. R. Rao Satellite Centre, Bangalore.}

\begin{abstract}
 We present an X-ray study of the high-synchrotron-peaked blazar Mrk~421 using ten days of XPoSat/XSPECT observations (2025 January 20--29; 202~ks) and contemporaneous NICER data. The 0.8--15~keV light curve has a fractional variability amplitude of $19.6 \pm 0.3$ per cent, and Bayesian blocks resolve 26 flux states spanning a factor of 2.3. The best-constrained flare has a fast rise and a slow decay, with e-folding times of about 3~h and 30~h. For typical jet parameters, the X-ray-emitting electrons cool faster than this, suggesting that the decay traces a gradual weakening of their acceleration. The count-rate distribution is best described by two lognormal components, consistent with multiplicative variability. The power spectrum is a red-noise power law with a slope of about 1.7 over timescales of 25~min to 10~d, typical of blazars. XSPECT and NICER count rates are well correlated over eight matched epochs ($r = 0.941$). We fitted log-parabola models to day-resolved spectra from XSPECT, NICER and both jointly, and to flux-resolved XSPECT spectra. Both show spectral hardening with brightness. In the flux-resolved spectra, the photon index at 1~keV decreases from 2.24 in the faintest state to 1.96 in the brightest, and the synchrotron peak moves from about 0.66 to 1.06~keV, while the curvature ($\beta \approx 0.7$) shows no significant change. Power-law fits to Bayesian blocks show that bright blocks at the same rate can differ in photon index by up to $6.4\sigma$, so count rate alone does not fix the spectral state.
\end{abstract}


\section{Introduction}\label{sec:intro}

Blazars are a subclass of radio-loud active galactic nuclei (AGN) whose
relativistic jets are oriented close to the observer's line of sight, so that
Doppler boosting dominates their non-thermal emission and renders them
exceptionally luminous and variable \citep{1995PASP..107..803U,
2003ApJ...596..847B}. They exhibit strong, stochastic flux variability across
the entire electromagnetic spectrum, on time-scales ranging from minutes to
years \citep{1997ARA&A..35..445U, bhatta2020nature, bhatta2016multifrequency}.
Their broadband spectral energy distribution (SED) has a characteristic
two-hump structure \citep{Fossati1998, ghisellini1997optical}: the low-energy
hump, extending from radio to X-rays, is synchrotron emission from relativistic
electrons in the jet, while the high-energy hump, peaking in the $\gamma$-rays,
is commonly attributed to inverse-Compton scattering in leptonic models,
through the synchrotron self-Compton \citep[SSC;][]{maraschi1992jet,
1993ApJ...407...65G} or external-Compton \citep[EC;][]{1992A&A...256L..27D,
sikora1994comptonization, 2000ApJ...545..107B} processes, with hadronic
scenarios also proposed \citep{mannheim1993proton, mucke2001proton}. On the
basis of the synchrotron-peak frequency, BL~Lac objects are divided into low-,
intermediate-, and high-synchrotron-peaked sources (LBL, IBL, HBL), the last
with $\nu_{p,\mathrm{syn}} > 10^{15}$~Hz \citep{abdo2010spectral}.

Markarian~421 (Mrk~421), located at
RA~$=11^{\mathrm{h}}04^{\mathrm{m}}27.31^{\mathrm{s}}$ and
Dec~$=+38^{\circ}12'31.8''$ (J2000), is the prototypical HBL and, at a redshift
of $z = 0.031$. It was detected initially as a
weak $\gamma$-ray source by EGRET \citep{lin1992detection} and shortly later become  among the nearest and brightest very-high-energy (VHE) blazars
in the extragalactic sky, first detected at TeV energy with the Whipple
telescope \citep{punch1992detection, petry1996detection}. 
It was the first extragalactic source established as a TeV emitter.  Its proximity minimises attenuation by the extragalactic background light (EBL), allowing tight
constraints on its intrinsic VHE spectrum \citep{Aharonian2006, Mazin2007}.
 
The source has been the target of numerous multiwavelength campaigns with
both space- and ground-based facilities \citep{shukla2012multiwavelength,
sinha2016long, krawczynski2001simultaneous}, and is persistently bright in the
optical, X-ray, and TeV bands, frequently showing rapid ($\lesssim 1$--2~day)
variability in the X-ray and TeV regimes \citep{blazejowski2005, Aleksic2015,
Balokovi2016, Markowitz2022}. This rapid X-ray variability implies that the
radiating electrons are highly energetic with short cooling times
\citep{2007ApJ...664L..71A, ackermann2016minute}. Observations with the Large
High Altitude Air Shower Observatory (LHAASO) have detected Mrk~421 up to PeV
energies, placing it among the most extreme particle accelerators known
\citep{cao2024first, abeysekara2020great}. During the major 2010 outburst,
strong X-ray/VHE correlations were recorded, and VERITAS observed a historic
flare reaching $\sim 27$~Crab above 1~TeV \citep{Aleksic2015,
abeysekara2020great, abe2025characterization}. In the optical, long-term and
intraday monitoring has revealed significant variability together with a clear
\textit{bluer-when-brighter} trend and strong, variable polarization, including
EVPA rotations that point to an evolving jet magnetic-field structure
\citep{Tosti1998, Gaidos1996, Carnerero2017, Singh2019}. More recently, X-ray
polarimetry with IXPE has opened a new window on the jet magnetic field:
\citet{Kim2024} found a consistently high X-ray polarization degree ($\sim
14\%$) with intermittent variations in the polarization angle, including
rotations of the angle relative to the jet axis.
 
In both flaring and quiescent states, the X-ray spectrum of Mrk~421 is
characteristically curved and is well described by a log-parabola model
\citep{fossati2000x, 2004A&A...413..489M, tramacere2007signatures}. Because the
synchrotron peak of an HBL lies in the soft X-ray band, this regime directly
probes the highest-energy electrons and their acceleration and cooling, so that
flux variability is typically accompanied by pronounced spectral evolution
\citep{tramacere2007signatures, massaro2004log}. A range of physical
models---broken power law, log-parabola, and power law with a maximum electron
energy---have been fitted to this curvature, and correlations among their
parameters have been exploited to break spectral degeneracies and to isolate the
drivers of variability \citep{2021MNRAS.508.5921H, 10.1093/mnras/stac1964,
2025JHEAp..45..438A}. In a multi-epoch AstroSat study, \citet{2025JHEAp..45..438A} found the
X-ray pivot energy to be constant across epochs and independent of the adopted
spectral model, indicating that the variability is driven chiefly by changes in
the spectral index rather than the normalisation, and pointing to fluctuations
in the particle acceleration or escape time-scales as the dominant cause.
Building on this, \citet{2026JHEAp..5300633A} showed that Mrk~421 exhibits a
two-state spectral behaviour that emerges independently from flux-resolved
 and  time-resolved spectroscopy. The time resolved analysis showed that the flux distribution is double log-normal, establishing two state characteristic is as an
intrinsic property of the source rather than an artefact of a particular
analysis. The log-normal form of this flux distribution provides a further diagnostic, it  suggests a multiplicative variability, and has been reported
in several blazars \citep{2005MNRAS.359..345U, 2010A&A...520A..83H,
pankaj_ln}.

These AstroSat studies were carried out with pointed observations, which
provide limited data, so that only a handful of epochs are available for
time-resolved spectral analysis. The diagnostics used in such studies could
be tested more robustly with soft X-ray monitoring that is densely sampled
and continuous over many days.
The X-ray Spectroscopy and Timing instrument (XSPECT) on board XPoSat,
designed for long uninterrupted monitoring of bright sources, is well suited
to this task. In this paper, we combine XSPECT with contemporaneous NICER
observations to exploit the
complementary soft-band coverage and timing of the two instruments. We present a detailed time-domain X-ray spectral analysis of Mrk~421 from
a joint XPoSat/XSPECT and NICER campaign. We characterise the source's
variability using Bayesian-block segmentation, perform day-resolved and
flux-resolved spectroscopy to track the spectral evolution, test whether the
X-ray count rate uniquely determines the spectral state, and compare the two
instruments. The paper is organised as follows.
Section~\ref{sec:obs} describes the XSPECT and NICER data and their reduction;
Section~\ref{sec:var} presents the variability analysis;
Section~\ref{sec:cross} the cross-instrument comparison;
Section~\ref{sec:spec} the spectral analysis;
Section~\ref{sec:disc} discusses the implications; and
Section~\ref{sec:summary} summarises our conclusions.

\section{Observations and Data Reduction}
\label{sec:obs}

\subsection{XPoSat XSPECT}
\label{sec:obs_xspect}

XSPECT is a collimated soft X-ray spectrometer aboard XPoSat, built around Swept
Charge Devices \citep[SCDs;][]{2001NIMPA.458..568L, 2008SPIE.7021E..17H}, a
detector technology previously flown on Chandrayaan-1
\citep[C1XS;][]{2009P&SS...57..717G}, Chandrayaan-2
\citep[CLASS;][]{2011LPI....42.1708R} and \textit{Insight}-HXMT
\citep{2020SCPMA..6349505C}. It comprises 16 SCDs, of which 15 are open and one
is covered by a 500~$\mu$m tantalum sheet to monitor the in-orbit particle
background \citep{rkrish2025, 2025arXiv250609918C}. The open detectors feed two
collimators with fields of view of approximately $2^\circ \times 2^\circ$
(detectors 0--3 and 8--11) and $3^\circ \times 3^\circ$ (detectors 4--7 and
12--14). The instrument covers 0.8--15~keV with an energy resolution of about 180~eV at
6~keV \citep{2026arXiv260721365N} and a timing resolution of 1~ms.

Mrk~421 was observed on ten consecutive days from 2025 January 20 to 29 under
proposal T25\_0002. Level-2 products were retrieved from PRADAN and the ten
daily screened event files were combined with the \texttt{xspevtmerge} utility,
which merges the event lists together with their good time interval extensions.
The merged file contains 261 intervals totalling 202.0~ks.

Transient particle events were screened with \texttt{xspfilterflares}, which
identifies intervals in which the count rate above the 16~keV threshold used by
the tool exceeds 225 counts per 10~s across the eleven monitored detectors, and
removes those intervals with 15~s of padding either side. No
exposure was removed, the good time interval structure being identical before
and after filtering, consistent with the saturated count rate remaining flat
and well below threshold throughout.

Detector-wise light curves were generated to check for the localised
contamination described in the instrument documentation, in which one or more
detectors of quadrant~1 can register spurious counts near the day-night
terminator. Detector~4 showed a series of discrete excursions reaching 15 to
67~count~s$^{-1}$ against a baseline of approximately 0.5~count~s$^{-1}$, while
detectors 5, 6 and 7 agreed with detectors 12, 13 and 14 to within a few per
cent. Detector~4 was therefore excluded from all subsequent products, leaving
eight detectors for the $2^\circ \times 2^\circ$ field of view and six for the
$3^\circ \times 3^\circ$. Because this departs from the default detector
selection, the ancillary response and background files supplied with the
instrument calibration are not applicable, and both were regenerated for the
adopted selection.

Only single events were used, this being the only event selection for which
instrument responses and background models are defined. Light curves were
extracted over 0.8--15~keV at 10~s binning for each field of view, together
with the matching background light curves. The background was subtracted from
each bin, with the uncertainties combined in quadrature. The light curve was
then rebinned to 500~s, treating each contiguous stretch of data separately so
that no 500~s bin spans an orbital gap. A stretch shorter than 500~s, or the
leftover part of a longer stretch that does not divide evenly into 500~s,
gives a bin with less than 500~s of exposure. We kept only bins with at least
250~s of exposure, half of the nominal 500~s; the rest were discarded,
leaving 404 bins. The mean background-subtracted rate is 1.188~count~s$^{-1}$ in the
$2^\circ \times 2^\circ$ field and 1.284~count~s$^{-1}$ in the
$3^\circ \times 3^\circ$.

Spectra were extracted from the same merged event file with matching ancillary
responses and background spectra. For all the spectral fits presented in
Section~\ref{sec:spec} the background was estimated with the static
prescription, in which both the cosmic X-ray background and the galactic
cosmic-ray template are held at their blank-sky averages. Over the co-added day-wise and flux-resolved
exposures used here this is appropriate, since the source counts exceed the
background below 6~keV. The hybrid prescription, in which the galactic
cosmic-ray template is rescaled to the particle level measured during the
observation, is used only for the per-block fixed-rate test of
Section~\ref{sec:spec_rate}, where the short individual blocks would otherwise
be over-subtracted; as shown there, the choice does not affect that result.

\begin{table}[htbp]
\centering
\caption{Log of the \textit{NICER} observations of Mrk~421 used in this work.
MJD is the observation start; Exp.\ is the good-time exposure after screening.
ObsID 7566043201 had zero good time after screening and is not listed.}
\label{tab:obs-details}
\begin{tabular}{lccc}
\hline
ObsID & Day & MJD & Exp.\ (s) \\
\hline
7566042901 & 01 & 60695.429 & 368 \\
7566043001 & 02 & 60696.264 & 488 \\
7566043101 & 03 & 60697.049 & 151 \\
7100110149 & 05 & 60699.892 & 172 \\
7100110150 & 06 & 60700.015 & 253 \\
7566043301 & 06 & 60700.011 & 319 \\
7566043401 & 07 & 60701.049 & 247 \\
7566043501 & 08 & 60702.017 & 122 \\
7566043601 & 09 & 60703.164 & 495 \\
\hline
\end{tabular}
\\[2pt]\raggedright\footnotesize Total exposure 2.6~ks; individual exposures 122--495~s.
\end{table}
\subsection{NICER}
\label{sec:obs_nicer}

The Neutron Star Interior Composition Explorer \citep[NICER;][]{Gendreau2016} is
a soft X-ray timing instrument aboard the International Space Station. Its X-ray
Timing Instrument comprises 56 co-aligned pairs of X-ray concentrator optics and
silicon drift detectors, of which 52 are currently operational
\citep{Malacaria2023}, providing photon-counting spectroscopy over 0.2--12~keV
with a timing precision of about 100~ns. The soft-band sensitivity of NICER is
well suited to a high synchrotron peaked blazar such as Mrk~421, whose
synchrotron emission extends into the X-ray band.

Ten NICER observations of Mrk~421 were obtained between MJD 60695 and 60703,
overlapping the XSPECT campaign (Table~\ref{tab:obs-details}). The data were
reduced with HEASoft v6.34 and the NICER calibration release CALDB 20240206,
using XSPEC v12.14.1 for the subsequent spectral analysis. The \texttt{nicerl2}
pipeline was run to produce calibrated, screened event files, applying the
standard good time interval criteria that exclude passages of the South Atlantic
Anomaly, low Earth-limb elevation, high optical loading and intervals of
enhanced particle background. One observation, 7566043201, has zero good time
after screening and was discarded, leaving nine pointings with individual
exposures between 122 and 495~s and a combined exposure of 2.6~ks.

Spectral products, including the source spectrum and the associated ARF and RMF
response files, were generated with \texttt{nicerl3-spect}. Background spectra
were produced with the SCORPEON model. Unlike the library-based 3C50 and Space
Weather models, SCORPEON is a parameterised model that describes both the
non-X-ray background (trapped particles, cosmic rays and South Atlantic Anomaly
contributions) and the astrophysical X-ray background (the cosmic X-ray
background, the Galactic halo and the local hot bubble) with smooth templates
that can be fitted jointly with the source model, giving a closer match to the
data. The source spectra were grouped with the optimal binning scheme of
\citet{2016A&A...587A.151K} (\texttt{grouptype optmin}) with a minimum of 25
counts per group. The spectra were fitted over 0.4--8~keV, where the calibration
is well established and the source dominates the background. Light curves were
extracted with \texttt{nicerl3-lc} at 100~s binning; the SCORPEON background
contributes 0.3 per cent of the source rate, so the choice of background model
has no material effect on the light curve. The mean background-subtracted rate
over the merged campaign is $509.5 \pm 3.5$~count~s$^{-1}$ in 0.3--10~keV.

\section{Variability}
\label{sec:var}

\begin{figure*}[tbp]
\centering
\includegraphics[width=\textwidth]{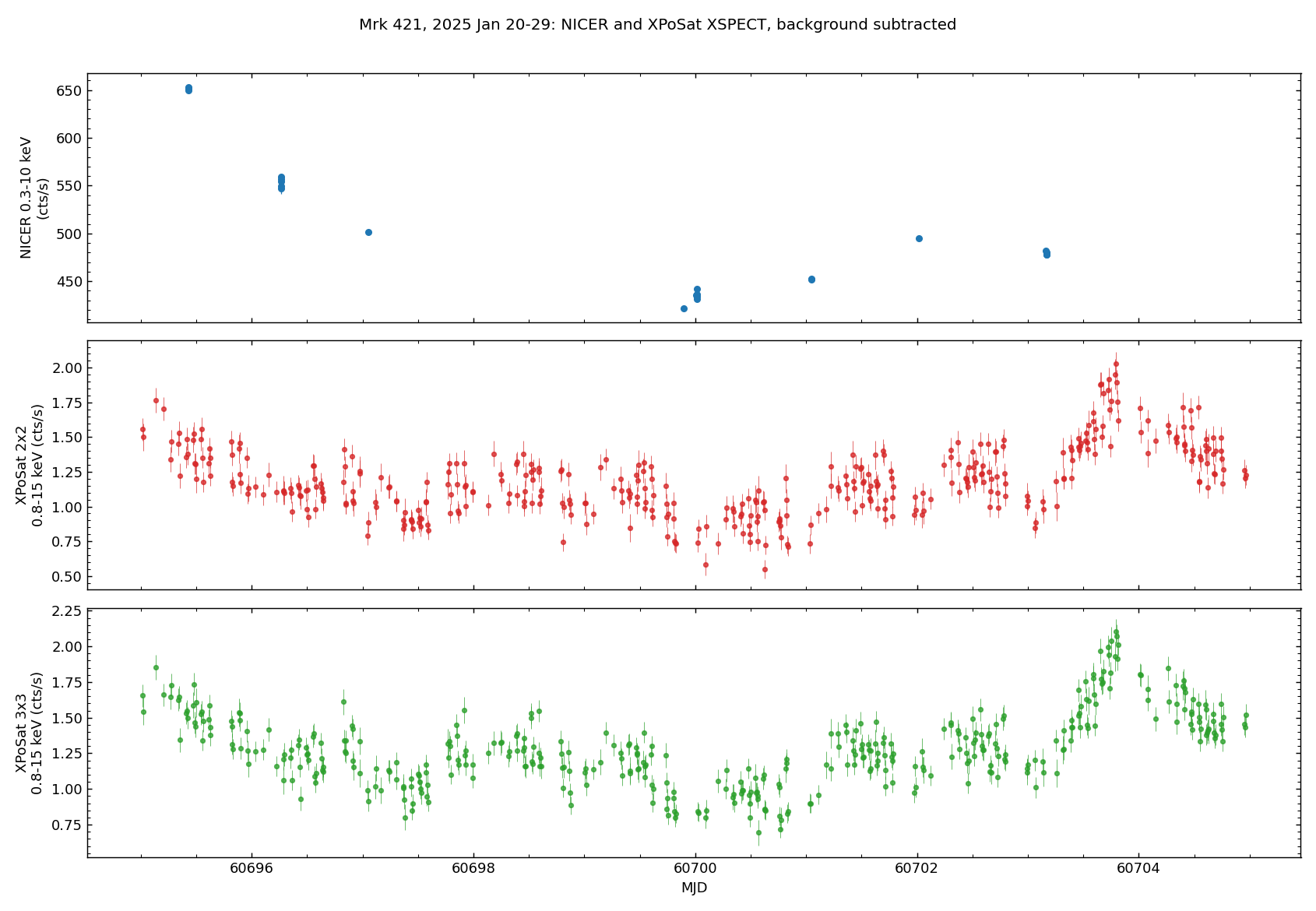}
\caption{Background-subtracted light curves of Mrk 421 over MJD 60695--60705.
Top: NICER 0.3--10~keV. Middle and bottom: XSPECT 0.8--15~keV for the
$2^\circ \times 2^\circ$ and $3^\circ \times 3^\circ$ fields of view, binned at
500~s.}
\label{fig:lc}
\end{figure*}

The background-subtracted XSPECT and NICER light curves over the full
campaign are shown in Figure~\ref{fig:lc}.

\subsection{Fractional variability}
\label{sec:var_fvar}

The fractional variability amplitude ($F_{\rm var}$) was computed following
\citet{Vaughan2003}. Over the 404 good 500~s bins of the
$2^\circ \times 2^\circ$ light curve the mean rate is 1.188~count~s$^{-1}$,
the total variance is $6.07 \times 10^{-2}$ and the mean square measurement
error $6.42 \times 10^{-3}$, so that the excess variance accounts for 89 per
cent of the total. The resulting amplitude is
$F_{\rm var} = 19.6 \pm 0.3$ per cent. The 500~s bin rates range from 0.548 to
2.028~count~s$^{-1}$ over the campaign.

For comparison, \citet{Markowitz2022} report $F_{\rm var}$ between 8.6 and
15.2 per cent for Mrk 421 in the AstroSat SXT sub-bands over a four-day
baseline. Blazar X-ray variability follows a red-noise power spectrum, whose
power rises steeply towards longer timescales, so a longer monitoring baseline
samples more of this low-frequency variance and tends to return a larger
$F_{\rm var}$ even for an intrinsically unchanged variability process. The
larger amplitude found here, over a ten-day rather than a four-day baseline,
is therefore consistent with this effect and does not by itself indicate that
Mrk~421 was intrinsically more variable during this campaign.

\subsection{Bayesian block decomposition}
\label{sec:var_blocks}

The background-subtracted, 500~s-binned light curve was segmented using the
Bayesian blocks algorithm \citep{Scargle2013}, as implemented in
\textsc{astropy}. This algorithm uses the point measures fitness function to
group together consecutive points whose count rates are statistically
consistent with each other. Each such group, or ``block,'' represents a period
during which the source flux stayed constant within the measurement errors.

The XSPECT observation has a duty cycle of only 23 per cent. This means the
source was actually observed for only about a quarter of the total time. The
rest is mostly gaps caused by the spacecraft's orbit. These gaps happen at
regular intervals. Because of this, a change-point algorithm like Bayesian
blocks could mistake the start or end of a gap for a real change in the
source's count rate.
To check this, we simulated 200 light curves of a constant source. Each one
used the same time sampling and the same gaps as the real XSPECT data. Since
the simulated source does not vary, any blocks found in these simulations must
come from the gaps or from random noise, not from real changes in brightness.
We repeated this test for different values of the false-alarm parameter $p_0$.
On average, the simulations gave only about 1--2 blocks, and in no realisation at any threshold did the number of blocks approach that found in the data.
So the many blocks seen in the real light curve are not caused by the orbital
gaps or by chance alone.
As expected, the number of blocks found in the real data depends on $p_0$. It
drops from 48 blocks at $p_0 = 0.3$ to 17 blocks at $p_0 = 0.001$. We use
$p_0 = 0.05$, which gives 27 blocks. The exact number and position of the
blocks change with $p_0$. But the main variability features discussed below
are seen at every value of $p_0$ we tested. 

\begin{figure*}[tbp]
\centering
\includegraphics[width=\textwidth]{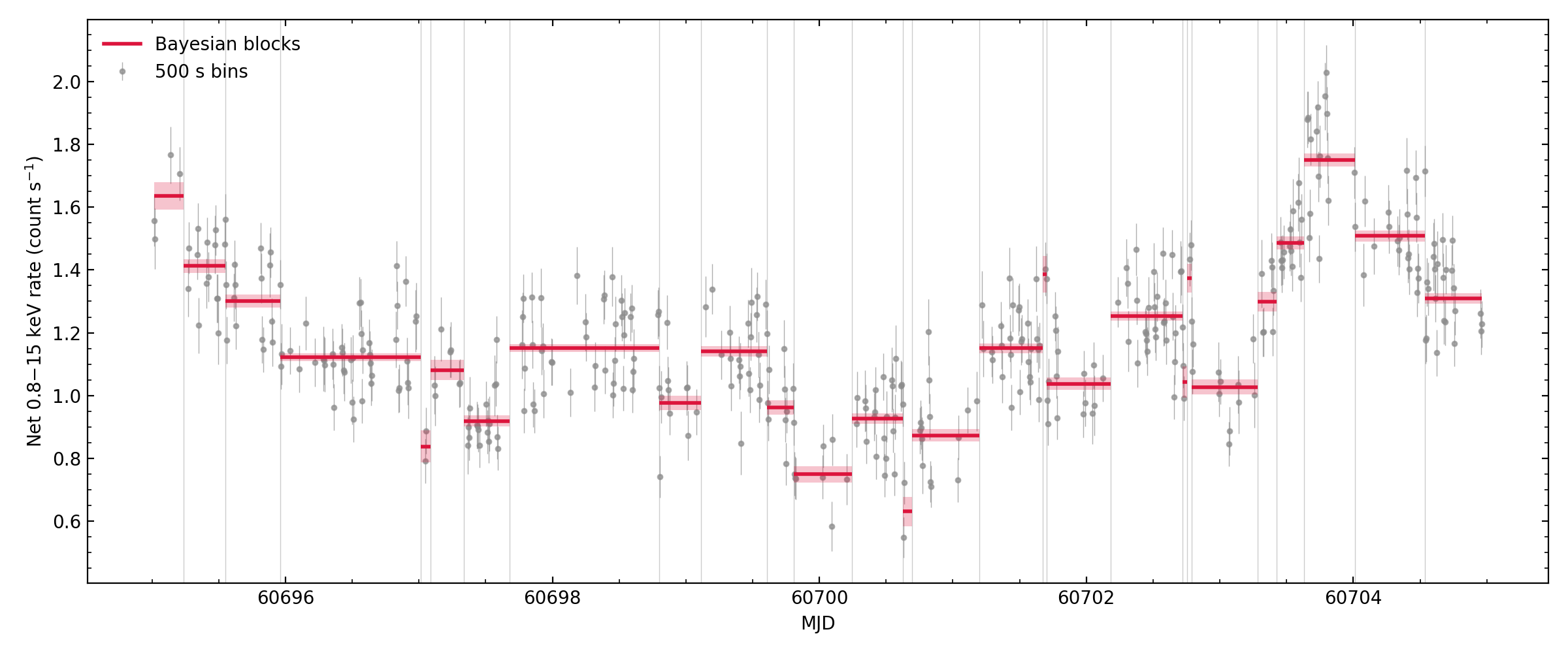}
\caption{XSPECT $2^\circ \times 2^\circ$ light curve at 500~s resolution (grey)
with the Bayesian block representation overlaid (red), for
$p_0 = 0.05$. Shaded regions indicate the uncertainty on each block rate.}
\label{fig:blocks}
\end{figure*}

One of these blocks, block 14, is excluded as instrumental
(Section~\ref{sec:var_fov}), leaving 26 blocks whose properties are listed in
Table~\ref{tab:blocks} and whose representation is shown in
Figure~\ref{fig:blocks}. The faintest, block 12, has a rate of
$0.749 \pm 0.026$~count~s$^{-1}$ at MJD 60699.811--60700.242, and the
brightest, block 25, $1.751 \pm 0.021$~count~s$^{-1}$ at MJD
60703.633--60704.012, a ratio of 2.34 significant at $30\sigma$. Of the 25
transitions between adjacent retained blocks, 24 are significant at more than
$3.8\sigma$. The one exception is the join between blocks 13 and 15. These two
blocks sit on either side of the excluded block 14. They differ by only
$2.1\sigma$, which is not a real difference. In other words, blocks 13 and 15
are consistent with the same flux level. The only reason the algorithm treated
them as separate blocks is the excised interval sitting between them.

In addition to these block-to-block changes, the light curve also shows a
broader pattern that plays out gradually over the ten-day
campaign: the source declines from $1.635 \pm 0.044$~count~s$^{-1}$ at the
start to a minimum around MJD 60700, rises to its highest point near MJD
60703.8, and declines again over the final day. Four of the blocks are much
shorter than the rest, lasting only 2.75 to 6.25~ks; these are brief dips or
rises from the surrounding level, superimposed on this broader pattern.

\subsection{Constraints on transition timescales}
\label{sec:var_tau}

We examined the 25 boundaries identified by the Bayesian-block analysis to
check whether the corresponding changes in count rate were actually observed,
or happened during an orbital gap. One boundary is the join between blocks 13
and 15. This boundary exists only because block 14, which originally sat
between them, was removed as instrumental; once it is excised, blocks 13 and
15 become directly adjacent. The count rates of blocks 13 and 15 differ by
only $2.1\sigma$, so this is not a real change in the source, and we do not
count it among the resolved transitions.
Of the remaining 24 boundaries, 14 fall within orbital gaps lasting 3.5 to
15.5~ks. Because the source was not observed during these gaps, we cannot pin
down when the transition happened -- only that it occurred somewhere within the
gap.
The other 10 boundaries fall during continuous observations, so each can be
located to within a single 500~s bin. All 10 of these transitions are observed fall during the
decrease in count rate. This does not mean that the source declines more
often than it rises. Locating a transition precisely
requires continuous coverage spanning it, and by chance the continuous stretches of data in this campaign happen to fall on declining parts of the
light curve.

The most rapid apparent change among the well-localised transitions occurred at
MJD 60700.627, but this was found to be instrumental
(Section~\ref{sec:var_fov}). Once we exclude
it, the remaining transition takes 500~s or longer. This is simply the
finest time resolution our data allow; we cannot see anything faster.
 If a source brightens or fades
within some time $\tau$, the emitting region making that light cannot be bigger than $\sim c\tau$, the distance light travels in that time: light from the far side of a larger region arrives later than light from the near side. In our case, however, the timescale we can measure is set by our own 500~s bins, not by the source; the source may well vary faster than 500~s, and our data simply cannot show it. Any size limit we quoted would therefore  only reflect our bin size, not a real property of the source, so we do not attempt to derive one.

\subsection{Cross-collimator consistency check}
\label{sec:var_fov}
The two fields of view provide an internal consistency
check, since they view the source through co-aligned colimators with completely separate sets of detectors. Over the 404
good bins, the ratio between the $2^\circ \times 2^\circ$ and $3^\circ \times 3^\circ$ count rates has a mean of 0.929. The scatter of these ratios is rms $\sim$  0.095  is close to what we expect from counting statistics alone (0.086), so
the two fields of view agree well overall; the small remaining scatter is only
about 4 per cent.

A single bin, at MJD 60700.6295, departs from the mean
ratio by more than 3$\sigma$, with a ratio of 0.640. To examine this stand out, we looked at the light curves of
each individual detector during this bin. All eight $2^\circ \times 2^\circ$
detectors drop together, by 1 to 27 per cent, with an average drop of 16 per
cent. All seven $3^\circ \times 3^\circ$ detectors, however, stay constant to
within 6 per cent.
A real change in the source's brightness, such as a flare or a dip, would show
up in both fields of view together, since they are looking at the same source
at the same time. Here the drop appears in every detector of one field of view
and in none of the other, so it cannot be a real change in the source. It is
more likely instrumental, for example, a small shift in the spacecraft's
pointing, to which the narrower $2^\circ \times 2^\circ$ field of view would
respond more strongly. For this reason, we exclude the Bayesian block built
around this bin (block 14) from the variability analysis. No other interval in
the campaign shows a similar departure.

\begin{table*}[tbp]
\centering
\caption{Bayesian block decomposition of the XSPECT $2^\circ \times 2^\circ$
light curve, $p_0 = 0.05$. Block 14 is excluded as instrumental
(Section~\ref{sec:var_fov}).}
\label{tab:blocks}
\begin{tabular}{rccrrr}
\hline
Block & MJD start & MJD stop & Span (ks) & Exposure (ks) & Rate (count s$^{-1}$) \\
\hline
1  & 60695.0161 & 60695.2389 & 19.25 &  2.38 & $1.635 \pm 0.044$ \\
2  & 60695.2389 & 60695.5514 & 27.00 &  7.54 & $1.412 \pm 0.021$ \\
3  & 60695.5514 & 60695.9623 & 35.50 &  7.72 & $1.301 \pm 0.020$ \\
4  & 60695.9623 & 60697.0126 & 90.75 & 20.12 & $1.122 \pm 0.012$ \\
5  & 60697.0126 & 60697.0849 &  6.25 &  1.00 & $0.838 \pm 0.051$ \\
6  & 60697.0849 & 60697.3367 & 21.75 &  3.40 & $1.081 \pm 0.032$ \\
7  & 60697.3367 & 60697.6810 & 29.75 &  8.58 & $0.918 \pm 0.018$ \\
8  & 60697.6810 & 60698.7979 & 96.50 & 20.79 & $1.151 \pm 0.012$ \\
9  & 60698.7979 & 60699.1133 & 27.25 &  5.59 & $0.977 \pm 0.023$ \\
10 & 60699.1133 & 60699.6081 & 42.75 & 11.10 & $1.141 \pm 0.017$ \\
11 & 60699.6081 & 60699.8106 & 17.50 &  4.90 & $0.962 \pm 0.023$ \\
12 & 60699.8106 & 60700.2418 & 37.25 &  4.36 & $0.749 \pm 0.026$ \\
13 & 60700.2418 & 60700.6266 & 33.25 & 10.88 & $0.927 \pm 0.016$ \\
14 & 60700.6266 & 60700.6932 &  5.75 &  1.09 & excluded \\
15 & 60700.6932 & 60701.1995 & 43.75 &  7.08 & $0.873 \pm 0.020$ \\
16 & 60701.1995 & 60701.6712 & 40.75 & 13.14 & $1.151 \pm 0.015$ \\
17 & 60701.6712 & 60701.7030 &  2.75 &  0.93 & $1.387 \pm 0.058$ \\
18 & 60701.7030 & 60702.1804 & 41.25 &  7.24 & $1.038 \pm 0.020$ \\
19 & 60702.1804 & 60702.7215 & 46.75 & 15.38 & $1.254 \pm 0.014$ \\
20 & 60702.7215 & 60702.7533 &  2.75 &  1.20 & $1.043 \pm 0.050$ \\
21 & 60702.7533 & 60702.7910 &  3.25 &  1.49 & $1.374 \pm 0.046$ \\
22 & 60702.7910 & 60703.2858 & 42.75 &  4.84 & $1.027 \pm 0.024$ \\
23 & 60703.2858 & 60703.4275 & 12.25 &  3.37 & $1.298 \pm 0.031$ \\
24 & 60703.4275 & 60703.6330 & 17.75 &  7.64 & $1.487 \pm 0.021$ \\
25 & 60703.6330 & 60704.0120 & 32.75 &  8.17 & $1.751 \pm 0.021$ \\
26 & 60704.0120 & 60704.5386 & 45.50 & 10.52 & $1.508 \pm 0.019$ \\
27 & 60704.5386 & 60704.9640 & 36.75 & 11.07 & $1.309 \pm 0.016$ \\
\hline
\end{tabular}
\end{table*}

\section{Cross-Instrument Comparison}
\label{sec:cross}

\subsection{Epoch matching}
\label{sec:cross_match}
NICER observed the source in nine separate pointings. We group any pointings separated by less than 0.02~d ($\approx$29~minutes) into a single visit. This reduces the nine pointings to eight visits: two of them, both on MJD 60700, are close enough in time to merge into one.
Strictly simultaneous coverage between NICER and XSPECT is minimal: only two
of the eight visits overlap in time with an XSPECT good time interval at all.
Summed over those two visits, the true overlap totals just 254~s, containing
roughly 350 net XSPECT counts. This is insufficient for a spectral
measurement and marginal for a flux measurement. The scarcity of overlap
follows from NICER pointings averaging only 290~s, XSPECT's own duty cycle
being reduced to 23 per cent by Earth occultation and South Atlantic Anomaly
passage, and the two spacecraft having uncorrelated visibility windows.
Each NICER visit was matched to the nearest qualifying XSPECT good time
interval. A good time interval here is a single, continuous, uninterrupted
stretch of XSPECT exposure from one orbital pass. Its length is generally
different from the 500~s bins used elsewhere in the light-curve analysis.
Only good time intervals of at least 400~s duration were treated as
qualifying, to avoid matching a NICER visit to a brief fragment with too few counts
for a reliable measurement. If the nearest good time interval to a visit is
shorter than 400~s, it is skipped, and the next-nearest qualifying interval is
used instead. Seven of the eight visits are matched within one hour, and five
within thirty minutes.  Within the matched intervals, the XSPECT count rate is
stable to between 0.6 and 14 per cent. This compares with 9 to 20 per cent
variation over a full day, so each matched interval samples a reasonably
well-defined flux state. Individual matching offsets are approximate at the
few-minute level, since the light curves are binned at 500~s. The matched
epochs are listed in Table~\ref{tab:epochs}.

\begin{table}[htbp]
\centering
\caption{Matched XSPECT and NICER epochs. $\Delta t$ is the interval between
the NICER visit and the nearest qualifying XSPECT good time interval.}
\label{tab:epochs}
\begin{tabular}{rcrrrr}
\hline
Epoch & MJD & $\Delta t$ & Exp. & $R_{\rm XSPECT}$ & $R_{\rm NICER}$ \\
 & & (min) & (ks) & (s$^{-1}$) & (s$^{-1}$) \\
\hline
1 & 60695.431 &  9 & 1.80 & 1.408 & 651.6 \\
2 & 60696.266 & 25 & 1.13 & 1.104 & 553.2 \\
3 & 60697.050 &  0 & 0.83 & 0.840 & 501.1 \\
4 & 60699.893 & 95 & 2.33 & 0.832 & 421.3 \\
5 & 60700.015 &  3 & 1.17 & 0.790 & 435.7 \\
6 & 60701.051 & 13 & 0.88 & 0.799 & 452.1 \\
7 & 60702.018 & 37 & 1.47 & 0.996 & 494.6 \\
8 & 60703.167 & 37 & 0.57 & 1.007 & 479.6 \\
\hline
\end{tabular}
\end{table}

\subsection{Flux correlation}
\label{sec:cross_corr}
\begin{figure}[htbp]
\centering
\includegraphics[width=\columnwidth]{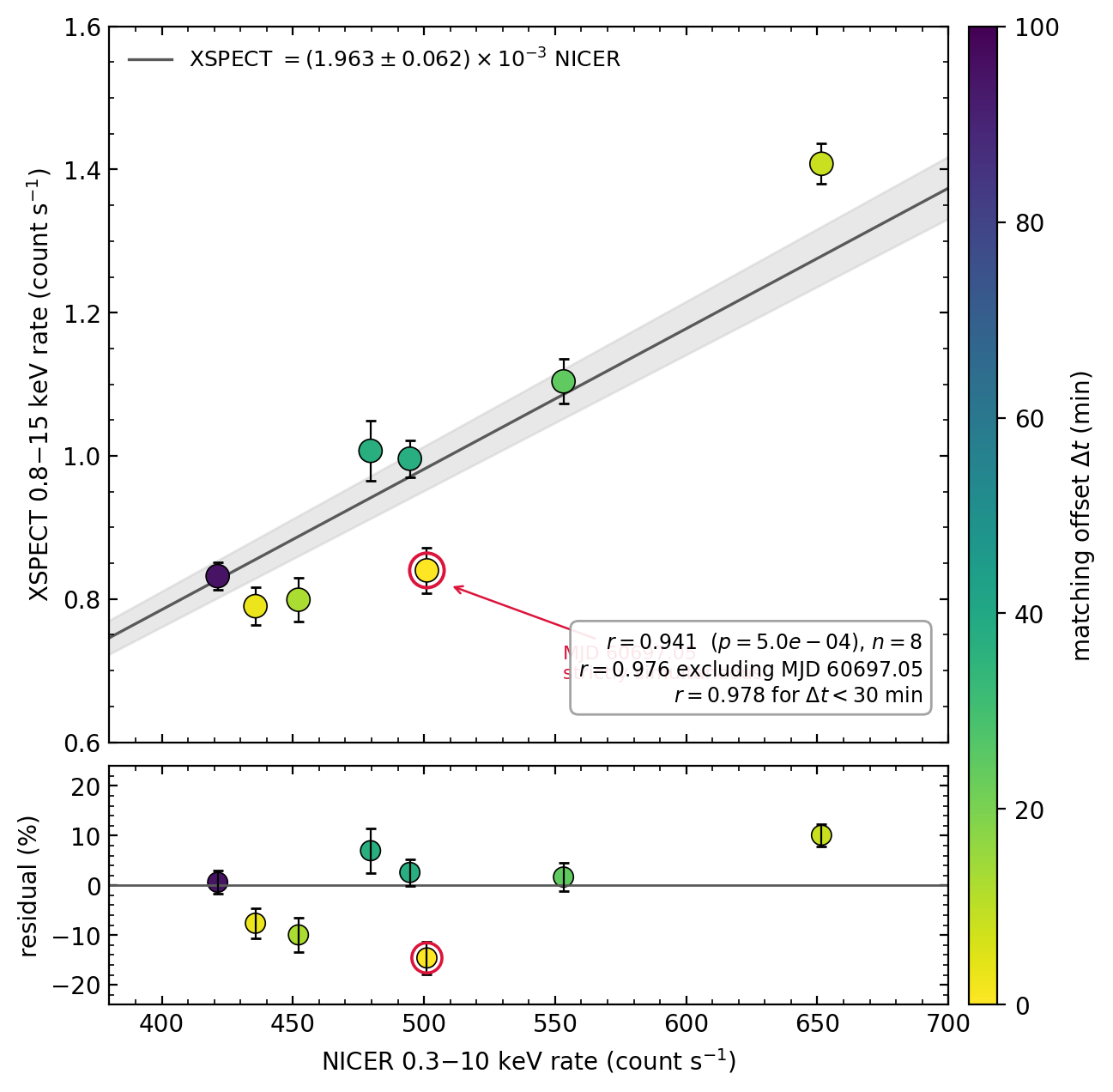}
\caption{XSPECT against NICER count rate for the eight matched epochs, coloured
by matching offset. The line and shaded band show the proportional fit and its
$1\sigma$ uncertainty. The epoch at MJD 60697.05, the only one with strictly
simultaneous coverage and the only significant outlier, is ringed. Lower panel:
fractional residuals about the fit.}
\label{fig:corr}
\end{figure}

Over the eight matched epochs, the NICER count rate spans 421 to 652~count~s$^{-1}$, and the XSPECT count rate spans 0.79 to 1.41~count~s$^{-1}$. The two rates are well correlated (see correlation Figure ~\ref{fig:corr}): the Pearson coefficient is 0.941 ($p = 5 \times 10^{-4}$), and the Spearman coefficient is 0.833 ($p = 0.010$). The scatter is dominated by a single epoch, at MJD 60697.05, which is discussed further in Section~\ref{sec:cross_ep3}. Excluding that epoch, the correlation rises to 0.976 ($p = 1.7 \times 10^{-4}$, $n = 7$). Restricting to the five epochs matched within thirty minutes gives 0.978 ($p = 0.004$). However, the anomalous epoch has a nominal timing offset of zero, so it survives every such cut. This improvement likely reflects the smaller sample size as much as the tighter time matching.

A linear fit to the eight epochs gives an intercept of
$-0.35 \pm 0.20$~count~s$^{-1}$. This is only 1.8 times its own uncertainty,
well short of the roughly 3$\sigma$ usually needed to call a result
significant, so the data are consistent with a zero intercept. Additionally, the eight epochs span only a narrow range of flux, from
421 to 652~count~s$^{-1}$ in NICER, a factor of 1.55. None of the data lie
anywhere near zero, so reading off the intercept means extending the fitted
line far past the range where we actually have measurements -- a small change
in slope, well within the noise, would move that crossing point
substantially. For both reasons, we instead assume the two rates are simply
proportional and fit a line through the origin:
\begin{equation}
\begin{split}
R_{\rm XSPECT}(0.8\text{--}15\ {\rm keV}) = {} & (1.963 \pm 0.062)
\times 10^{-3}\, \\
& \times R_{\rm NICER}(0.3\text{--}10\ {\rm keV}),
\end{split}
\end{equation}
equivalent to a rate ratio of $510 \pm 16$.
The uncertainty ($\pm 16$) describes how precisely we know the average ratio
across the whole campaign; it is not the uncertainty on any single epoch's
individual measurement. For that, we compare each epoch's actual XSPECT rate
to what the fitted relation predicts from its NICER rate: on average, epochs
differ from the relation by 8.5 per cent. We therefore adopt this 8.5 per
cent as the uncertainty when converting one
individual measurement from one instrument's rate to the other.
This scatter of 8.5 percent can be explained by three known effects: Poisson counting noise
on the XSPECT rates, 2 to 4 per cent; real changes in the source's brightness
during the gap between the NICER visit and its matched XSPECT interval; and the
size of that timing gap itself for each epoch (Section~\ref{sec:cross_match}),
which controls how much room the source has to vary in between.

 We also checked whether this NICER/XSPEC count rate ratio changes with the source's overall
brightness, but find no significant trend ($r = -0.48$,
$p = 0.23$).
The trend we do measure is negative, which is the sign expected if the source hardens as it brightens, as found directly from the spectral fits (Section~\ref{sec:spec}). NICER is
sensitive down to 0.3~keV, while XSPECT only sees above 0.8~keV, so NICER
picks up low-energy photons that XSPECT misses entirely. If the source hardens as it brightens then XSPECT's rate
should rise slightly faster than NICER's as the source brightens. This would
pull the NICER-to-XSPECT ratio down with increasing flux, which is exactly
the negative trend seen here. Because the measured trend is not statistically
significant and with only eight epochs, this comparison offers only weak, independent support for the hardening when source brightens.

\subsection{The MJD 60697.05 epoch}
\label{sec:cross_ep3}

This epoch has the most strictly simultaneous NICER and XSPECT coverage of any
in the campaign, at 151~s, and is the only significant outlier in the flux
correlation, its XSPECT rate lying 14.6 per cent below the fitted relation. The
Bayesian block decomposition isolates a short low state at this time, with a
rate of $0.838 \pm 0.051$~count~s$^{-1}$ over 1.0~ks of exposure, against
$1.122 \pm 0.012$ in the preceding block and $1.081 \pm 0.032$ in the following
one, decreases significant at $5.4\sigma$ and $4.0\sigma$ respectively. The
ratio of the two XSPECT fields of view over this block is 0.88, within the
normal scatter (Section~\ref{sec:var_fov}), so the low state is seen by both
collimators and is not instrumental.

We first tested whether the deficit could be spectral rather than a drop in
total luminosity. NICER is sensitive down to 0.3~keV, whereas XSPECT sees only
above 0.8~keV, so a softening of the spectrum that redistributed flux from
higher to lower energies at roughly constant output would depress the XSPECT
rate while leaving the NICER rate unchanged or slightly raised. We fitted the
NICER spectrum of the pointing coincident with this epoch (ObsID 7566043101,
151~s) and the campaign-averaged NICER spectrum simultaneously, with the
Galactic column shared and fixed at $N_{\rm H} = 1.34 \times 10^{20}$~cm$^{-2}$,
using a log-parabola in which the photon index, curvature and normalisation were
free in each spectrum. The coincident pointing gives $\alpha = 2.354$
(2.338--2.370) and $\beta = 0.259$ (0.208--0.309), against $\alpha = 2.342$
(2.334--2.350) and $\beta = 0.274$ (0.255--0.293) for the campaign average, at
90 per cent confidence. The difference in photon index is $0.012 \pm 0.018$,
consistent with zero. The coincident pointing is part of the campaign average
(about 6 per cent of its exposure), so the two spectra are not strictly
independent, but this small overlap would if anything reduce a real difference
and does not affect the conclusion. The coincident NICER rate of
$501.1 \pm 5.0$~count~s$^{-1}$ also shows no rise relative to the campaign
average of $509.5 \pm 3.5$. We therefore find no evidence that the deficit is
spectral in origin.

The explanation is instead one of sampling. The 151~s NICER pointing covers
only 15 per cent of the 1.0~ks XSPECT block that spans the low state. Within
this simultaneous overlap the XSPECT rate is $0.779 \pm 0.131$~count~s$^{-1}$,
close to the block low-state rate of $0.838 \pm 0.051$ and only $1.8\sigma$
below the daily mean of $1.019 \pm 0.011$. This overlap measurement is too
imprecise to confirm the low state on its own, but it is fully consistent with
the state persisting and gives no indication that it had ended. The XSPECT
block, defined to span the low state, records the dip directly, whereas the
NICER matched rate averages predominantly over exposure outside the block, when
the source was at its normal brightness, so the dip barely registers in the
NICER average. No disagreement between the two instruments at any single
instant is required: the deficit is what results when a brief, real event is
caught by one instrument's continuous coverage and largely missed by the
other's short, offset pointing.

This is a general limitation of comparing a variable source across matched,
non-simultaneous intervals, and it cannot be removed here, strictly
simultaneous data amounting to only 254~s across the campaign at a per-interval
precision of order 20 per cent.

\subsection{Flux Distributions}
\begin{figure}[ht]
\centering
\includegraphics[width=\linewidth]{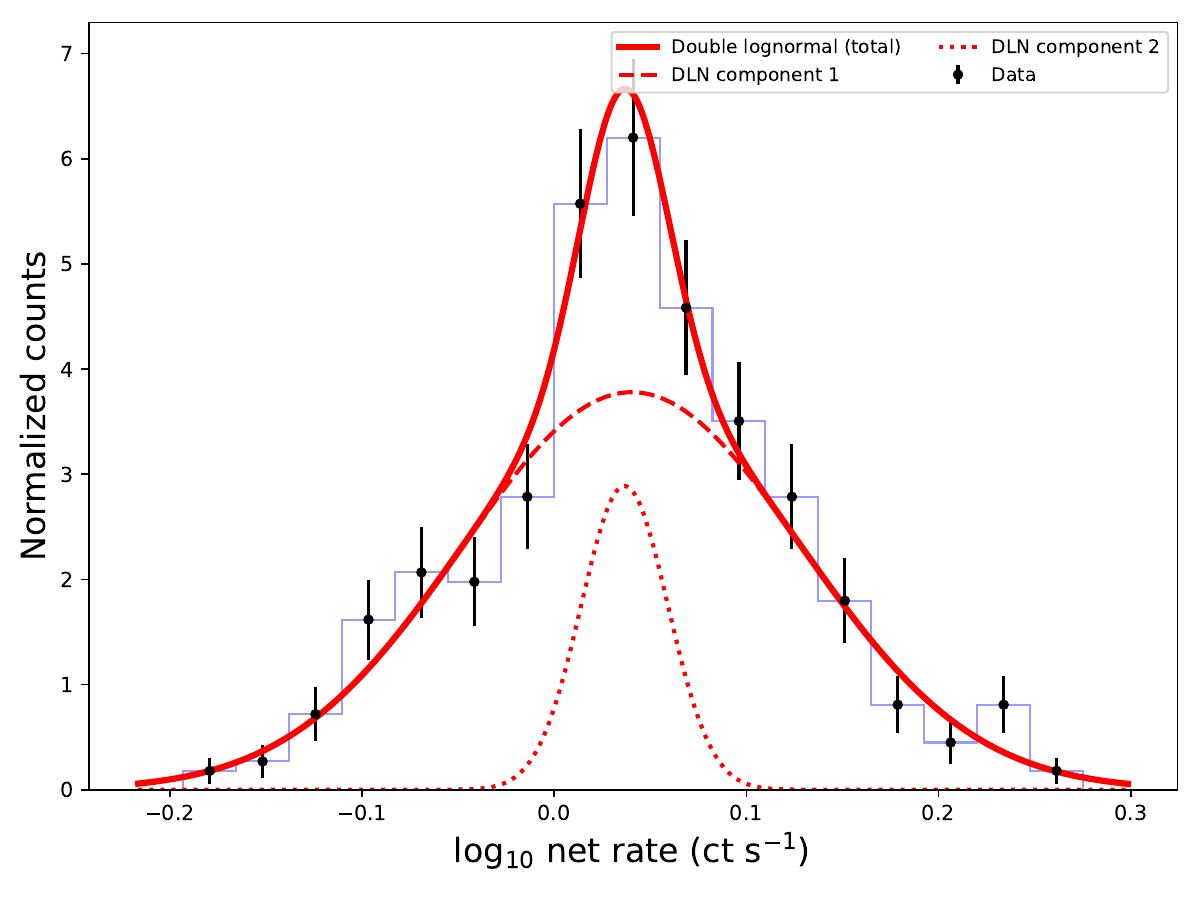}
\caption{Normalized histogram of $\log_{10}$ of the 0.8--7\,keV net count rate (ct\,s$^{-1}$) for FOV2,
with the double lognormal fit (solid) and its two components (dashed and dotted). Error bars are Poisson.}
\label{fig:xspect_dist}
\end{figure}

To investigate the statistical behaviour of the variability during the XSPECT observation of Mrk~421,
we examined the distribution of the 500\,s net count rate in the 0.8--7\,keV band for the
$2^\circ\times2^\circ$ field of view (FOV2), using the
Anderson--Darling (AD) test together with histogram fitting. Both the single normal and the single
lognormal models were unable to represent the observed distribution, as indicated by AD statistics that exceed
the 5\% critical value (0.779) and the 1\% critical value (1.081) for 404 points
(Table~\ref{tab:xspect_dist}). To obtain a more accurate description of the probability density function (PDF), we
constructed a normalized histogram of the logarithm of the count rate with 17 bins (Figure~\ref{fig:xspect_dist})
and fitted it using two-component models, namely the double lognormal and double normal functions defined in
Eqs.~\ref{eq:xspect_double_lognorm}--\ref{eq:xspect_double_norm}.

\begin{equation}
\label{eq:xspect_double_lognorm}
\begin{split}
D_{\mathrm{dLN}}(x)
&= a\,\frac{1}{\sqrt{2\pi}\,\sigma_{1}}
   \exp\!\left[-\frac{(x-\mu_{1})^{2}}{2\sigma_{1}^{2}}\right]  \\
&\quad + (1-a)\,\frac{1}{\sqrt{2\pi}\,\sigma_{2}}
   \exp\!\left[-\frac{(x-\mu_{2})^{2}}{2\sigma_{2}^{2}}\right],
\end{split}
\end{equation}
where $x=\log_{10}F$, $a$ is the mixing fraction, $\mu_1$ and $\mu_2$ are the centroids of the two components, and
$\sigma_1$ and $\sigma_2$ are their corresponding widths.

\begin{equation}
\label{eq:xspect_double_norm}
D_{\rm dN}(x) =
a\,N(x;\sigma_{1},\mu_{1}) + (1-a)\,N(x;\sigma_{2},\mu_{2}),
\end{equation}
with
\begin{equation}
N(x;\sigma,\mu) =
\frac{1}{\sqrt{2\pi}\,\sigma}
\exp\!\left[-\frac{(10^{x}-\mu)^{2}}{2\sigma^{2}}\right]
\,10^{x}\ln 10 .
\end{equation}

The AD statistics and the fit parameters are given in Table~\ref{tab:xspect_dist}. The double lognormal model
gives a reduced chi-square of $\chi^{2}_{\nu}=1.06$, compared with 1.69 for the single lognormal, 1.91 for the single
normal and 1.55 for the double normal. The count-rate
distribution is therefore better described by two lognormal components than by a single lognormal or by the
normal models.

\begin{table*}
\centering
\scriptsize
\setlength{\tabcolsep}{3pt}
\caption{Distribution statistics for the 0.8--7\,keV net count rate $F$ (ct\,s$^{-1}$) of the XSPECT observation (FOV2).
$A^2_{\log F}$ and $A^2_F$ are the AD statistics of $\log_{10}F$ (lognormal test) and of $F$ (normal test); the
5\% and 1\% critical values are 0.779 and 1.081. $\chi^2_{1}$ and $\chi^2_{2}$ are the reduced chi-square of the
single and the double lognormal, and $\chi^2_{\rm dN}$ that of the double normal. The double lognormal parameters
(Eq.~\ref{eq:xspect_double_lognorm}) are fitted to a 17-bin histogram of $\log_{10}F$. The errors are those of the fit.}
\label{tab:xspect_dist}
\begin{tabular}{lcccccccccc}
\toprule
FOV & $A^2_{\log F}$ & $A^2_F$ & $\chi^2_{1}$ & $\chi^2_{2}$ & $\chi^2_{\rm dN}$ & $a$ & $\mu_1$ & $\sigma_1$ & $\mu_2$ & $\sigma_2$ \\
\midrule
FOV2 & 1.10 & 2.83 & 1.69 & 1.06 & 1.55 & $0.84\pm0.07$ & $0.040\pm0.005$ & $0.089\pm0.005$ & $0.037\pm0.006$ & $0.023\pm0.007$ \\
\bottomrule
\end{tabular}
\end{table*}

\subsection{Flare Profile Fitting}
\label{sec:soe_xspect}
To characterize the flares in the 0.8--7\,keV light curve, we fitted the 500\,s net count rate with a sum of
exponentials (SOE) on a constant baseline,
\begin{equation}
\label{eq:soe_xspect}
F(t) = F_{b} + \sum_{k=1}^{N_p} \frac{2F_{p,k}}{\exp\!\left(\dfrac{t_{p,k}-t}{\tau_{r,k}}\right)
+ \exp\!\left(\dfrac{t-t_{p,k}}{\tau_{d,k}}\right)},
\end{equation}
where $F_b$ is the baseline, $F_{p,k}$ the amplitude, $t_{p,k}$ the time at which the rising and decaying exponentials
cross, and $\tau_{r,k}$ and $\tau_{d,k}$ the e-folding rise and decay times of the $k$th flare. The flux doubling and
halving times are $t_{\rm doub}=\ln 2\,\tau_r$ and $t_{\rm half}=\ln 2\,\tau_d$. The asymmetry of each flare is
\begin{equation}
\zeta = \frac{\tau_d-\tau_r}{\tau_d+\tau_r},
\end{equation}
with $|\zeta|<0.3$ symmetric, $0.3\le|\zeta|<0.7$ moderately asymmetric and $|\zeta|\ge0.7$ asymmetric. The time of
maximum is $t_{\rm max}=t_p+\frac{\tau_r\tau_d}{\tau_r+\tau_d}\ln(\tau_d/\tau_r)$.

We fitted five components, starting from the visible peaks, to the FOV2 light curve (MJD 60695.0--60705.0, 404 points).
The e-folding times were bounded between 1\,ks and 1.5\,d, and the baseline was bounded by the 5th
percentile of the rates. The fit was repeated from 31 starting points, and all converged to the same minimum. The
errors are the covariance errors scaled by $\sqrt{\chi^2_\nu}$. A component is accepted only if the value of each of its four
parameters exceeds twice its error and none lies at a bound.

\begin{figure*}[ht]
\centering
\includegraphics[width=\linewidth]{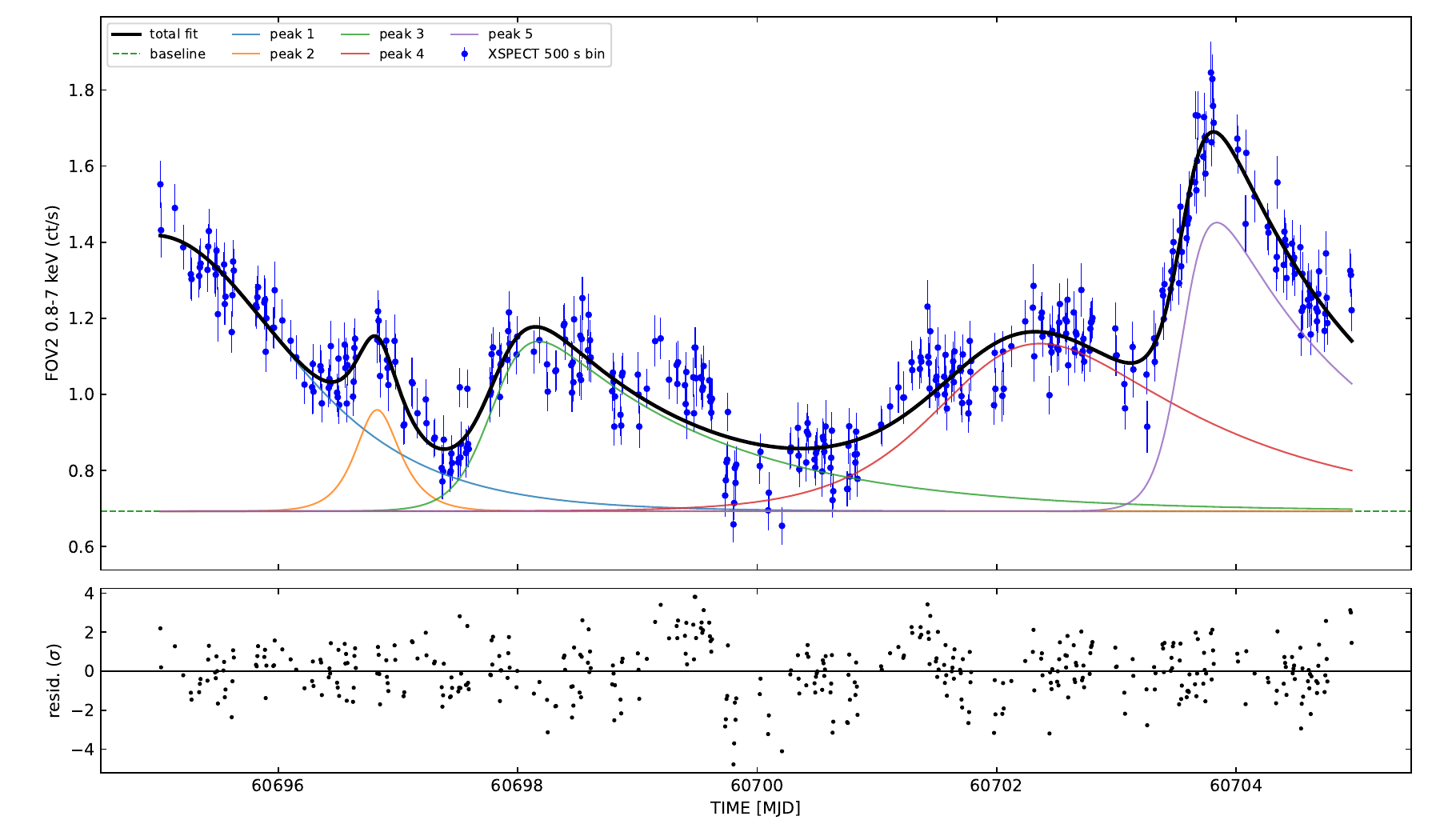}
\caption{The 0.8--7\,keV light curve (500\,s bins) of FOV2 with the five-component sum-of-exponentials fit
(Eq.~\ref{eq:soe_xspect}).}
\label{fig:soe_xspect}
\end{figure*}

The fit gives $\chi^2/{\rm NDF}=754.3/383$ ($\chi^2_\nu=1.97$), so the sum of five exponentials describes the
light curve but does not account for all of its variability
(Table~\ref{tab:soe_xspect}, Figure~\ref{fig:soe_xspect}). Removing any one component increases $\chi^2$ by at least
106, so each is needed to describe the light curve, but most rise and decay times are not constrained. Only one component
passes the acceptance criterion: the flare with maximum at MJD 60703.84, with
$\tau_r=0.124\pm0.020$\,d ($t_{\rm doub}=7.4$\,ks), $\tau_d=1.23\pm0.39$\,d  ($t_{\rm half}=73.7$\,ks)and $\zeta=0.82$, an asymmetric flare
with a fast rise and a slow decay. The decay times of the third and fourth flares are at the upper bound, the rise and decay
times of the second flare have errors comparable to their values, and those of the first flare, which is only partly
covered at the start of the observation, are not constrained by the data.

\begin{table*}
\centering
\footnotesize
\caption{Parameters of the sum-of-exponentials fit (Eq.~\ref{eq:soe_xspect}) to the 0.8--7\,keV light curve (FOV2). $t_p$ is
the crossing time and $t_{\rm max}$ the time of maximum (MJD), $F_p$ is in ct\,s$^{-1}$ and $\tau_r$, $\tau_d$ are in days.
Values marked with a dagger are at the upper bound (1.5\,d) and are lower limits. The baseline is
$0.69\pm0.18$\,ct\,s$^{-1}$. The component accepted by the criterion in the
text is marked with an asterisk.}
\label{tab:soe_xspect}
\begin{tabular}{lcccccccc}
\toprule
$k$ & $t_p$ & $F_p$ & $\tau_r$ & $\tau_d$ & $t_{\rm max}$ & $\zeta$ & $t_{\rm doub}$ (ks) & $t_{\rm half}$ (ks) \\
\midrule
1 & $60695.29\pm2.14$ & $0.69\pm0.59$ & $>1.5^{\dagger}$ & $0.80\pm1.08$ & 60694.96 & $-0.31$ & $>89.8$ & 47.7 \\
2 & $60696.82\pm0.13$ & $0.27\pm0.07$ & $0.16\pm0.10$ & $0.18\pm0.12$ & 60696.83 & 0.07 & 9.3 & 10.7 \\
3 & $60697.84\pm0.08$ & $0.31\pm0.06$ & $0.18\pm0.06$ & $>1.5^{\dagger}$ & 60698.18 & 0.79 & 10.5 & $>89.8$ \\
4 & $60701.96\pm0.31$ & $0.40\pm0.15$ & $0.57\pm0.22$ & $>1.5^{\dagger}$ & 60702.36 & 0.45 & 34.0 & $>89.8$ \\
5$^{*}$ & $60703.58\pm0.03$ & $0.52\pm0.06$ & $0.124\pm0.020$ & $1.23\pm0.39$ & 60703.84 & 0.82 & 7.4 & 73.9 \\
\bottomrule
\end{tabular}
\end{table*}

\subsection{Power Spectral Density}
\label{sec:psd_xspect}
To characterize the variability power spectrum of the 0.8--7\,keV light curve (FOV2, 404 points of 500\,s over
9.95\,d), we computed the Lomb--Scargle periodogram of the mean-subtracted rate at the Fourier frequencies $\nu_j = j/T$ ($j = 1, 2, \ldots$), where
$T = 9.95$\,d is the length of the light curve (MJD 60695.016--60704.964), and fitted a power law with a constant,
\begin{equation}
\label{eq:psd_pl}
P(\nu) = A\,\nu^{-s} + c .
\end{equation}
XSPECT observes the source for only about 1.5\,ks (up to three 500\,s bins) in each 5.83\,ks orbit, a duty cycle of
about 23\%. These regular gaps distort the periodogram: power leaks between frequencies and is aliased from near the
orbital frequency, and the periodogram values at neighbouring frequencies are no longer independent. Because of this,
we measured the slope in two ways: first by fitting the periodogram directly, and then by comparing it with simulated
light curves that have the same gaps as the real data.

First, we fitted Eq.~\ref{eq:psd_pl} to the periodogram below 7\,d$^{-1}$ (the alias-free limit for one sample per
orbit) by minimizing the Whittle statistic,
\begin{equation}
\label{eq:whittle}
S = 2\sum_{j}\left[\frac{I_j}{P(\nu_j)} + \ln P(\nu_j)\right],
\end{equation}
where $I_j$ is the periodogram at $\nu_j$, with $1\sigma$ errors from the profile likelihood ($\Delta S=1$).
This assumes independent ordinates, so the errors are lower limits.

Second, we used a simulation-based method in the style of PSRESP \citep{Uttley2002}. For each trial slope $s$
(PSD $\propto\nu^{-s}$, $0.5\le s\le3.5$ in steps of 0.05) we simulated 100 light curves
\citep{Timmer1995} that are longer than the observation, sampled them at exactly the observed times, rescaled
them to the observed mean and intrinsic variance, and added Gaussian noise of the mean measurement error. The binned periodogram of each
simulation was computed in 10 logarithmic frequency bins, and the observed periodogram was compared with the
simulations through
\begin{equation}
\label{eq:psresp}
\chi^2(s) = \sum_{i}\left[\frac{\ln P^{\rm obs}_i - \langle\ln P^{\rm sim}_i\rangle}{\sigma_i}\right]^2 ,
\end{equation}
where $\sigma_i$ is the standard deviation of $\ln P^{\rm sim}_i$ over the simulations. The best slope minimizes
$\chi^2(s)$. A slope is acceptable if at least 5\% of the simulations, evaluated against the same mean, have a
larger $\chi^2$ than the observation; we quote the contiguous range of acceptable slopes around the best one.

\begin{figure*}
\centering
\includegraphics[width=\linewidth]{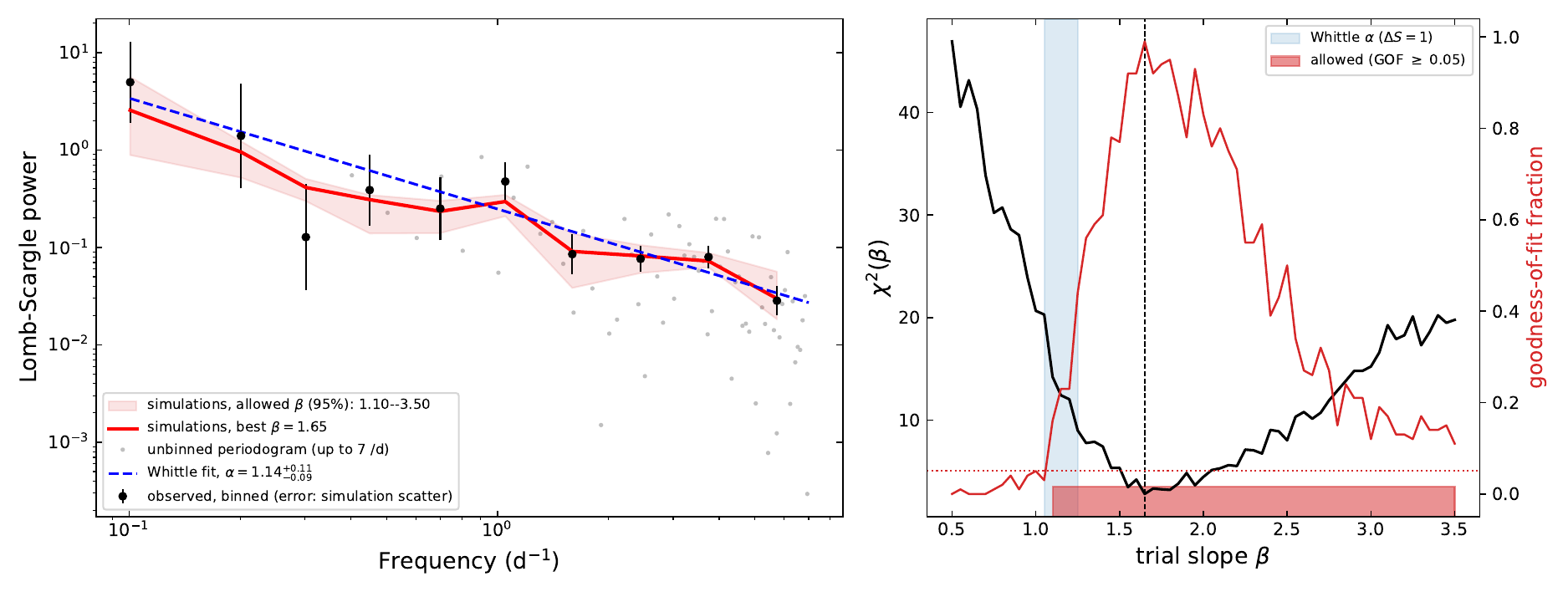}
\caption{Power spectrum of the 0.8--7\,keV light curve (FOV2). Left: binned Lomb--Scargle periodogram below
7\,d$^{-1}$ (points, with errors from the scatter of the simulations), unbinned periodogram (grey dots), best simulated
spectrum for $s=1.65$ (red line) with the range of the acceptable slopes (shaded) and the Whittle power-law fit
(dashed). Right: $\chi^2(s)$ (black) and the goodness-of-fit fraction (red, with the 0.05 level dotted); the
blue band marks the Whittle slope range.}
\label{fig:psd_xspect}
\end{figure*}

The results are listed in Table~\ref{tab:psd_xspect} and shown in Figure~\ref{fig:psd_xspect}. Using the periodogram
up to 57.6\,d$^{-1}$, the simulations give $s=1.70$ with an acceptable range of 1.25--3.05 and a goodness-of-fit
fraction of 0.93. The binned periodogram in this range shows a peak near 13\,d$^{-1}$, which the simulations reproduce;
it comes from the orbital sampling and is not intrinsic to the source. Restricting the simulations to frequencies below
7\,d$^{-1}$ gives $s=1.65$, with an acceptable range from 1.10 up to the edge of the grid (3.5) and a goodness-of-fit
fraction of 0.99. The Whittle fit in the same range gives $s=1.14^{+0.11}_{-0.09}$, inside the range of the simulations;
the constant $c$ is consistent with zero. Models with a bend in the power law do not improve the fit: the bend
frequency moves to the upper edge of the frequency range, so we find no evidence for a break. The data therefore
require a steep red-noise spectrum with $s\gtrsim1.1$, but the slope is poorly constrained, and we adopt
$s=1.7$ (range 1.25--3.05) from the simulations, which account for the gaps.

\begin{table}
\centering
\footnotesize
\setlength{\tabcolsep}{3pt}
\renewcommand{\arraystretch}{1.2}
\caption{Power spectral slope of the 0.8--7\,keV light curve (FOV2). The simulation range is the contiguous set of slopes
with a goodness-of-fit fraction of at least 0.05; for 0.1--7\,d$^{-1}$ it extends to the upper edge of the simulated grid ($s=3.5$), so only the lower bound is constrained. The Whittle error ($\Delta S=1$) assumes independent ordinates.}
\label{tab:psd_xspect}
\begin{tabular*}{\columnwidth}{@{\extracolsep{\fill}}lcccc@{}}
\toprule
Method & Freq. (d$^{-1}$) & Slope $s$ & Range & Fit frac. \\
\midrule
Simulations & 0.1--57.6 & $1.70$ & 1.25--3.05 & 0.93 \\
Simulations & 0.1--7    & $1.65$ & $\geq 1.10$ & 0.99 \\
Whittle fit & 0.1--7    & $1.14^{+0.11}_{-0.09}$ & 1.05--1.25 & -- \\
\bottomrule
\end{tabular*}
\end{table}

\section{Spectral Analysis}
\label{sec:spec}

The strong variability found in Section~\ref{sec:var} motivates a
detailed spectral analysis. We approach it at three levels: a day-resolved
analysis fitting each of the ten campaign days independently, both with XSPECT
alone and jointly with the contemporaneous NICER pointings; a flux-resolved
analysis co-adding the Bayesian blocks into three count-rate states; and a
test, using the Bayesian blocks, of whether the count rate uniquely determines
the spectral shape.

Throughout, spectra were extracted with the static background prescription. Over
the co-added exposures used here (each day-wise and flux-resolved spectrum
accumulates $\gtrsim 15$~ks) the source counts comfortably exceed the static
background below 6~keV, and the static estimate is that recommended by the
instrument team. The single exception is the per-block fixed-rate test of
Section~\ref{sec:spec_rate}, where the short individual blocks contain too few
counts for the static blank-sky rate to be subtracted without driving the net
rate negative; there we adopt the hybrid prescription, and we show below that
this does not affect the result because it rests on the power-law index, which
is insensitive to the background level.

The X-ray spectrum of Mrk~421 is well known to be curved rather than a
simple power law, consistent with a synchrotron peak lying within or near
the X-ray band; this curvature is commonly described by a log-parabola model
\citep{fossati2000x, massaro2004, tramacere2007signatures}. 
We adopt an absorbed log-parabola  ($\mathrm{TBabs} \times
\mathrm{logpar}$) as the baseline continuum model for the day-resolved, NICER-only, joint, and flux-resolved spectra, for which the signal-to-noise is sufficient to constrain the spectral curvature. The log-parabola model is defined as
\begin{equation}
N(E) = K \left(\frac{E}{E_1}\right)^{-\alpha - \beta \log_{10}(E/E_1)},
\end{equation}
with $E_1 = 1$~keV a fixed reference energy, $K$ the normalization,
$\alpha$ the photon index at $E_1$, and $\beta$ the curvature. We also tested an absorbed broken power-law model, but it did not provide a better description of the spectra than the log-parabola (Section \ref{sec:spec_fluxres}). The short spectra of the individual Bayesian blocks are treated separately in Section 5.3: because in most block the curvature cannot be constrained reliably, we use an absorbed power law only to obtain a uniform effective photon index for comparing blocks at the same count rate.
Fits are
performed over 0.8--6~keV, above which the background exceeds 70 per cent of
the counts in the $2^\circ \times 2^\circ$ field. The Galactic column was
fixed to
$N_{\rm H} = 1.34 \times 10^{20}$~cm$^{-2}$, the weighted-average
H\,\textsc{i} value toward Mrk~421 from the HI4PI survey \citep{HI4PI2016}.
Over this band the results are insensitive to this choice: adopting instead the
higher $1.9 \times 10^{20}$~cm$^{-2}$ of \citet{Kalberla2005} changes the
photon indices by less than 0.03. Leaving $N_{\rm H}$ free is not warranted,
since over this narrow band it is degenerate with the curvature. We adopt the $2^\circ \times 2^\circ$ field as our primary measurement
throughout. Its eight detectors do not include detector~4, so excluding that
detector (Section~\ref{sec:obs_xspect}) does not affect it. The
$3^\circ \times 3^\circ$ field did include detector~4, so its response had
to be rebuilt from the remaining six detectors, so we use the
$3^\circ \times 3^\circ$ field only as a consistency check, not as our
primary result.

A recurring narrow feature near 1.7~keV is present in the residuals of the
brighter spectra. It is well described by an absorption edge and, where the
edge energy is left free, converges on the Si~K edge (1.72--1.76~keV), an instrumental Si~K feature
associated with the silicon-based detector materials. Wherever this edge is
detected, we add it to the spectral model as an extra multiplicative
component, $\mathrm{edge} \times \mathrm{TBabs} \times \mathrm{logpar}$,
letting its energy and maximum optical depth $\tau_{\rm max}$ vary freely in
the fit. For spectra where the edge is not required, we fit without it and
note this explicitly.

\subsection{Day-resolved spectra}
\label{sec:spec_daywise}

Each campaign day was fitted independently. The XSPECT day-wise fits are
listed in Table~\ref{tab:xspect_daywise_edge}. The edge is significant on
several days, with $\tau_{\rm max}$ up to $\approx 0.22$ and an energy
consistent with Si~K; on the remaining days it is not required and the fit uses
$\mathrm{TBabs} \times \mathrm{logpar}$. Across the ten days the photon index
$\alpha$ correlates strongly and inversely with the 0.8--6~keV flux (Pearson
$r = -0.83$, $p = 0.003$; Table~\ref{tab:correlations}), the clearest
expression of the harder-when-brighter behaviour in the campaign, while the
curvature $\beta$ shows no significant dependence on flux.

\begin{table*}[tbp]
\centering
\caption{Day-wise XPoSat/XSPECT ($2^\circ\times2^\circ$, static background, 0.8--6~keV) log-parabola fits of Mrk~421. Where an instrumental absorption edge is detected, the model is \texttt{edge*TBabs*logpar} with the edge energy free and its parameters listed; otherwise \texttt{TBabs*logpar}. $N_{\rm H}$ frozen at $1.34\times10^{20}$~cm$^{-2}$; $F_{0.8-6}$ in $10^{-10}$~erg~cm$^{-2}$~s$^{-1}$. Uncertainties are 90 per cent confidence.}
\label{tab:xspect_daywise_edge}
\begin{tabular}{ccccccc}
\hline
Day & MJD & $\alpha$ & $\beta$ & $E_{\rm edge}$ (keV) & $\tau_{\rm max}$ & $F_{0.8-6}$ / $\chi^2_\nu$ \\
\hline
01 & 60695 & $2.041_{-0.111}^{+0.108}$ & $0.752_{-0.167}^{+0.169}$ & $1.734_{-0.091}^{+0.067}$ & $0.153_{-0.080}^{+0.082}$ & $4.72$ / $1.04$ \\
02 & 60696 & $2.166_{-0.092}^{+0.091}$ & $0.659_{-0.153}^{+0.156}$ & $\dots$ & $\dots$ & $3.85$ / $1.02$ \\
03 & 60697 & $2.090_{-0.130}^{+0.127}$ & $0.714_{-0.207}^{+0.216}$ & $1.484_{-0.144}^{+0.114}$ & $0.112_{-0.089}^{+0.093}$ & $3.43$ / $0.85$ \\
04 & 60698 & $2.110_{-0.129}^{+0.128}$ & $0.763_{-0.200}^{+0.203}$ & $1.767_{-0.120}^{+0.102}$ & $0.143_{-0.095}^{+0.096}$ & $3.83$ / $1.14$ \\
05 & 60699 & $2.225_{-0.098}^{+0.097}$ & $0.747_{-0.168}^{+0.173}$ & $\dots$ & $\dots$ & $3.44$ / $0.92$ \\
06 & 60700 & $2.239_{-0.136}^{+0.135}$ & $0.622_{-0.213}^{+0.217}$ & $1.764_{-0.079}^{+0.090}$ & $0.222_{-0.107}^{+0.110}$ & $2.90$ / $0.88$ \\
07 & 60701 & $2.116_{-0.088}^{+0.088}$ & $0.757_{-0.148}^{+0.152}$ & $\dots$ & $\dots$ & $3.72$ / $1.07$ \\
08 & 60702 & $1.960_{-0.113}^{+0.112}$ & $0.776_{-0.175}^{+0.178}$ & $1.599_{-0.068}^{+0.089}$ & $0.177_{-0.083}^{+0.085}$ & $4.15$ / $0.81$ \\
09 & 60703 & $1.987_{-0.073}^{+0.072}$ & $0.529_{-0.115}^{+0.116}$ & $\dots$ & $\dots$ & $5.18$ / $0.90$ \\
10 & 60704 & $1.974_{-0.102}^{+0.100}$ & $0.672_{-0.151}^{+0.154}$ & $1.751_{-0.078}^{+0.064}$ & $0.157_{-0.076}^{+0.077}$ & $4.85$ / $0.82$ \\
\hline
\end{tabular}
\end{table*}

The nine NICER pointings were fitted independently over 0.4--8~keV with the
SCORPEON background (Table~\ref{tab:nicer_perobsid}). The same edge is detected
in the brighter NICER exposures (days~01 and~02, at $\approx 3.4$ and
$5.4\sigma$ from the fit-statistic improvement) at a consistent energy, and in
both the log-parabola and broken-power-law parameterisations, confirming that
it is an instrumental feature rather than an artefact of the adopted continuum.
The NICER-only indices reproduce the harder-when-brighter trend
($r = -0.69$, $p = 0.042$).

\begin{table}[htbp]
\centering
\caption{\textit{NICER}-only per-ObsID log-parabola fits of Mrk~421 (\texttt{TBabs*logpar}, Scorpeon background, 0.4--8~keV, $N_{\rm H}$ frozen at $1.34\times10^{20}$~cm$^{-2}$). $F_{0.8-6}$ is the unabsorbed 0.8--6~keV flux in $10^{-10}$~erg~cm$^{-2}$~s$^{-1}$. An instrumental edge near 1.75--1.85~keV is significant on the brighter epochs (days~01, 02) and marginal or undetected on fainter ones. Uncertainties are 90 per cent confidence.}
\label{tab:nicer_perobsid}
\footnotesize
\setlength{\tabcolsep}{3pt}
\begin{tabular}{lccccc}
\hline
ObsID & Day & $\alpha$ & $\beta$ & $F_{0.8-6}$ & $\chi^2/$dof \\
\hline
7566042901 & 01 & $2.248_{-0.013}^{+0.013}$ & $0.328_{-0.031}^{+0.032}$ & $7.82$ & 77.2/97 \\
7566043001 & 02 & $2.318_{-0.012}^{+0.012}$ & $0.272_{-0.033}^{+0.033}$ & $6.48$ & 77.5/84 \\
7566043101 & 03 & $2.317_{-0.015}^{+0.015}$ & $0.266_{-0.047}^{+0.048}$ & $5.84$ & 78.6/79 \\
7100110149 & 05 & $2.392_{-0.016}^{+0.016}$ & $0.408_{-0.050}^{+0.051}$ & $4.61$ & 61.5/73 \\
7100110150 & 06a & $2.379_{-0.015}^{+0.015}$ & $0.333_{-0.045}^{+0.046}$ & $4.87$ & 79.8/82 \\
7566043301 & 06b & $2.389_{-0.014}^{+0.014}$ & $0.381_{-0.042}^{+0.043}$ & $4.78$ & 53.1/83 \\
7566043401 & 07$^{\dagger}$ & $2.283_{-0.015}^{+0.015}$ & $0.252_{-0.044}^{+0.044}$ & $5.37$ & 128.5/85 \\
7566043501 & 08 & $2.255_{-0.017}^{+0.017}$ & $0.367_{-0.051}^{+0.052}$ & $5.82$ & 54.4/72 \\
7566043601 & 09 & $2.208_{-0.011}^{+0.011}$ & $0.252_{-0.030}^{+0.030}$ & $5.96$ & 82.8/100 \\
\hline
\end{tabular}
\\[2pt]\raggedright\footnotesize $^{\dagger}$ Day 07 fits poorly ($\chi^2/$dof$=1.51$) in both log-parabola and broken-power-law models, independently of the joint fit.
\\[6pt]
\begin{tabular}{llcc}
\hline
\multicolumn{4}{c}{Instrumental edge (\texttt{edge}) where detected} \\
\hline
ObsID & Day & $E_{\rm edge}$~(keV) & $\tau_{\rm max}$ \\
\hline
7566042901 & 01 & $1.75_{-0.07}^{+0.15}$ & $0.071_{-0.031}^{+0.031}$ \\
7566043001 & 02 & $1.83_{-0.04}^{+0.05}$ & $0.108_{-0.033}^{+0.033}$ \\
7566043001 & 02$^{\rm BKN}$ & $1.85_{-0.07}^{+0.12}$ & $0.056_{-0.033}^{+0.034}$ \\
7566043101 & 03 & $1.91_{-0.11}^{+0.09}$ & $0.095_{-0.057}^{+0.059}$ \\
\hline
\end{tabular}
\end{table}

Finally, each day was fitted jointly, tying $\alpha$ and $\beta$ between the two
instruments and allowing a free cross-normalisation $C_{\rm NICER}$ (with
XSPECT fixed at unity). The joint fits are listed in
Table~\ref{tab:daywise_joint}. The XSPECT-only, NICER-only and joint indices
agree to better than 0.01 in $\alpha$ on every day, so that the joint fit does
not bias the spectral parameters. The cross-normalisation stays close to unity, $C_{\rm NICER} \approx 0.99$,
on days~01--03, when the source varied little within the day. It departs
further from unity on days with more intraday variability, reaching
$C_{\rm NICER} = 1.36$ on day~09, the final flare rise. This makes sense: a
brief NICER snapshot only needs to match the day-long XSPECT average when
the source's brightness stays roughly constant that day. When the source
varies rapidly within a day, a short snapshot can easily land on a moment
brighter or fainter than the day's average, so the two no longer agree as
closely. Day~06 was covered by two NICER pointings; fitted separately
against the same XSPECT day, they return consistent parameters
($\alpha = 2.381 \pm 0.014$ and $2.390 \pm 0.014$), an internal check on the
stability of the joint fit. The harder-when-brighter trend is present in the
joint fits ($r = -0.63$, $p = 0.10$), though less sharply than in the
XSPECT-only fits because the non-simultaneity scatters the tied index; the
correlations for all three approaches are collected in
Table~\ref{tab:correlations}.

\begin{table*}[tbp]
\centering
\caption{Day-wise joint \textit{NICER}+XPoSat/XSPECT spectral fits of Mrk~421. XSPECT ($2^\circ\times2^\circ$, static background, 0.8--6~keV) and \textit{NICER} (Scorpeon, 0.4--8~keV) are fitted jointly with the source parameters tied across the two instruments; $C_{\rm NICER}$ is the free cross-normalisation (XSPECT~$\equiv1$). $N_{\rm H}$ is frozen at $1.34\times10^{20}$~cm$^{-2}$. $F_{0.8-6}$ is the unabsorbed 0.8--6~keV flux in $10^{-10}$~erg~cm$^{-2}$~s$^{-1}$. Uncertainties are 90 per cent confidence. Day~06 has two \textit{NICER} exposures (ObsIDs 7100110150 and 7566043301), both listed; they give consistent parameters. An instrumental absorption edge near the Si~K energy was included for days~01 and~02, where it improved the fit; the edge energy and maximum optical depth are $E_{\rm edge}=1.744_{-0.059}^{+0.077}$~keV, $\tau_{\rm max}=0.065_{-0.028}^{+0.028}$ (day~01) and $E_{\rm edge}=1.840_{-0.048}^{+0.053}$~keV, $\tau_{\rm max}=0.093_{-0.030}^{+0.031}$ (day~02). On day~06 an edge is detected in both exposures: $E_{\rm edge}=1.818_{-0.060}^{+0.060}$~keV, $\tau_{\rm max}=0.063_{-0.026}^{+0.026}$ (ObsID 7100110150) and $E_{\rm edge}=1.768_{-0.063}^{+0.112}$~keV, $\tau_{\rm max}=0.074_{-0.037}^{+0.038}$ (ObsID 7566043301). The remaining days did not require an edge component.}
\label{tab:daywise_joint}
\begin{tabular}{ccccccc}
\hline
\multicolumn{7}{c}{\textit{Log-parabola} (\texttt{TBabs*logpar})} \\
\hline
Day & MJD & $\alpha$ & $\beta$ & $C_{\rm NICER}$ & $F_{0.8-6}$ & $\chi^2/{\rm dof}$ \\
\hline
01 & 60695 & $2.247_{-0.012}^{+0.012}$ & $0.363_{-0.028}^{+0.028}$ & $0.989\pm0.009$ & $7.71_{-0.05}^{+0.05}$ & 168.3/166 \\
02 & 60696 & $2.319_{-0.012}^{+0.012}$ & $0.288_{-0.029}^{+0.029}$ & $0.999\pm0.010$ & $6.46_{-0.04}^{+0.04}$ & 219.5/161 \\
03 & 60697 & $2.317_{-0.015}^{+0.015}$ & $0.310_{-0.039}^{+0.040}$ & $0.987\pm0.011$ & $5.73_{-0.06}^{+0.06}$ & 143.8/146 \\
05 & 60699 & $2.390_{-0.015}^{+0.015}$ & $0.440_{-0.041}^{+0.042}$ & $1.268\pm0.014$ & $4.59_{-0.05}^{+0.05}$ & 132.4/141 \\
06$^{a}$ & 60700 & $2.381_{-0.015}^{+0.014}$ & $0.365_{-0.041}^{+0.041}$ & $1.012\pm0.012$ & $4.84_{-0.04}^{+0.04}$ & 154.6/150 \\
06$^{b}$ & 60700 & $2.390_{-0.014}^{+0.014}$ & $0.393_{-0.038}^{+0.039}$ & $1.026\pm0.012$ & $4.82_{-0.04}^{+0.04}$ & 122.8/151 \\
07 & 60701 & $2.281_{-0.014}^{+0.014}$ & $0.356_{-0.037}^{+0.038}$ & $1.154\pm0.012$ & $5.28_{-0.05}^{+0.05}$ & 257.1/163 \\
08 & 60702 & $2.248_{-0.016}^{+0.016}$ & $0.358_{-0.039}^{+0.039}$ & $1.139\pm0.012$ & $5.84_{-0.06}^{+0.06}$ & 126.8/141 \\
09 & 60703 & $2.202_{-0.011}^{+0.011}$ & $0.243_{-0.026}^{+0.026}$ & $1.363\pm0.011$ & $5.98_{-0.04}^{+0.04}$ & 172.0/169 \\
\hline
\multicolumn{7}{c}{\textit{Broken power law} (\texttt{TBabs*bknpower})} \\
\hline
Day & MJD & $\Gamma_1$ & $E_{\rm break}$~(keV) & $\Gamma_2$ & $C_{\rm NICER}$ / $F_{0.8-6}$ & $\chi^2/{\rm dof}$ \\
\hline
01 & 60695 & $2.155_{-0.028}^{+0.023}$ & $1.28_{-0.10}^{+0.09}$ & $2.535_{-0.026}^{+0.026}$ & $0.987$ / $7.77$ & 198.2/167 \\
02 & 60696 & $2.221_{-0.044}^{+0.029}$ & $1.15_{-0.14}^{+0.11}$ & $2.539_{-0.028}^{+0.026}$ & $0.997$ / $6.44$ & 229.6/162 \\
03 & 60697 & $2.231_{-0.039}^{+0.028}$ & $1.33_{-0.19}^{+0.16}$ & $2.562_{-0.043}^{+0.042}$ & $0.985$ / $5.77$ & 149.5/145 \\
05 & 60699 & $2.313_{-0.038}^{+0.022}$ & $1.62_{-0.22}^{+0.11}$ & $2.826_{-0.082}^{+0.057}$ & $1.270$ / $4.60$ & 150.0/140 \\
06$^{a}$ & 60700 & $2.267_{-0.035}^{+0.029}$ & $1.235_{-0.111}^{+0.094}$ & $2.651_{-0.034}^{+0.034}$ & $1.009$ / $4.88$ & 149.0/149 \\
06$^{b}$ & 60700 & $2.274_{-0.032}^{+0.029}$ & $1.250_{-0.091}^{+0.104}$ & $2.681_{-0.031}^{+0.035}$ & $1.023$ / $4.82$ & 127.5/150 \\
07 & 60701 & $2.205_{-0.028}^{+0.024}$ & $1.42_{-0.12}^{+0.10}$ & $2.582_{-0.036}^{+0.035}$ & $1.154$ / $5.31$ & 264.8/162 \\
08 & 60702 & $2.175_{-0.027}^{+0.026}$ & $1.50_{-0.12}^{+0.14}$ & $2.572_{-0.039}^{+0.045}$ & $1.140$ / $5.87$ & 140.6/140 \\
09 & 60703 & $2.153_{-0.023}^{+0.020}$ & $1.47_{-0.15}^{+0.15}$ & $2.417_{-0.028}^{+0.030}$ & $1.364$ / $6.01$ & 183.9/168 \\
\hline
\end{tabular}
\\[2pt]
\raggedright
\footnotesize $^{a}$ Day~06 with \textit{NICER} ObsID 7100110150. \quad $^{b}$ Day~06 with \textit{NICER} ObsID 7566043301.
\end{table*}

\begin{table}[htbp]
\centering
\caption{Correlations among the log-parabola spectral parameters and the 0.8--6~keV flux for the three independent day-resolved fitting approaches: the \textit{NICER}-only per-ObsID fits (Table~\ref{tab:nicer_perobsid}), the XSPECT day-wise fits (Table~\ref{tab:xspect_daywise_edge}), and the joint \textit{NICER}+XSPECT day-wise fits (Table~\ref{tab:daywise_joint}). For each pair the Pearson $r$ and Spearman $\rho$ coefficients are given with their $p$-values. The harder-when-brighter trend ($\alpha$ decreasing with flux) is recovered in all three, most strongly in the XSPECT day-wise fits; the curvature $\beta$ shows no significant dependence on flux.}
\label{tab:correlations}
\footnotesize
\setlength{\tabcolsep}{2pt}
\begin{tabular}{llcc}
\hline
Sample & Pair & Pearson $r$ ($p$) & Spearman $\rho$ ($p$) \\
\hline
\textit{NICER}-only (N=9) & $\alpha$--$F$ & $-0.69$ (0.042) & $-0.75$ (0.020) \\
 & $\beta$--$F$ & $-0.38$ (0.317) & $-0.59$ (0.092) \\
 & $\alpha$--$\beta$ & $+0.56$ (0.116) & $+0.66$ (0.053) \\
\hline
XSPECT day-wise (N=10) & $\alpha$--$F$ & $-0.83$ (0.003) & $-0.76$ (0.011) \\
 & $\beta$--$F$ & $-0.29$ (0.423) & $-0.07$ (0.855) \\
 & $\alpha$--$\beta$ & $+0.04$ (0.914) & $-0.25$ (0.489) \\
\hline
Joint day-wise (N=8) & $\alpha$--$F$ & $-0.63$ (0.096) & $-0.69$ (0.058) \\
 & $\beta$--$F$ & $-0.39$ (0.344) & $-0.55$ (0.160) \\
 & $\alpha$--$\beta$ & $+0.60$ (0.115) & $+0.52$ (0.183) \\
\hline
\end{tabular}
\end{table}

\subsection{Flux-resolved spectra}
\label{sec:spec_fluxres}

\begin{figure*}[tbp]
\centering
\gridline{%
  \fig{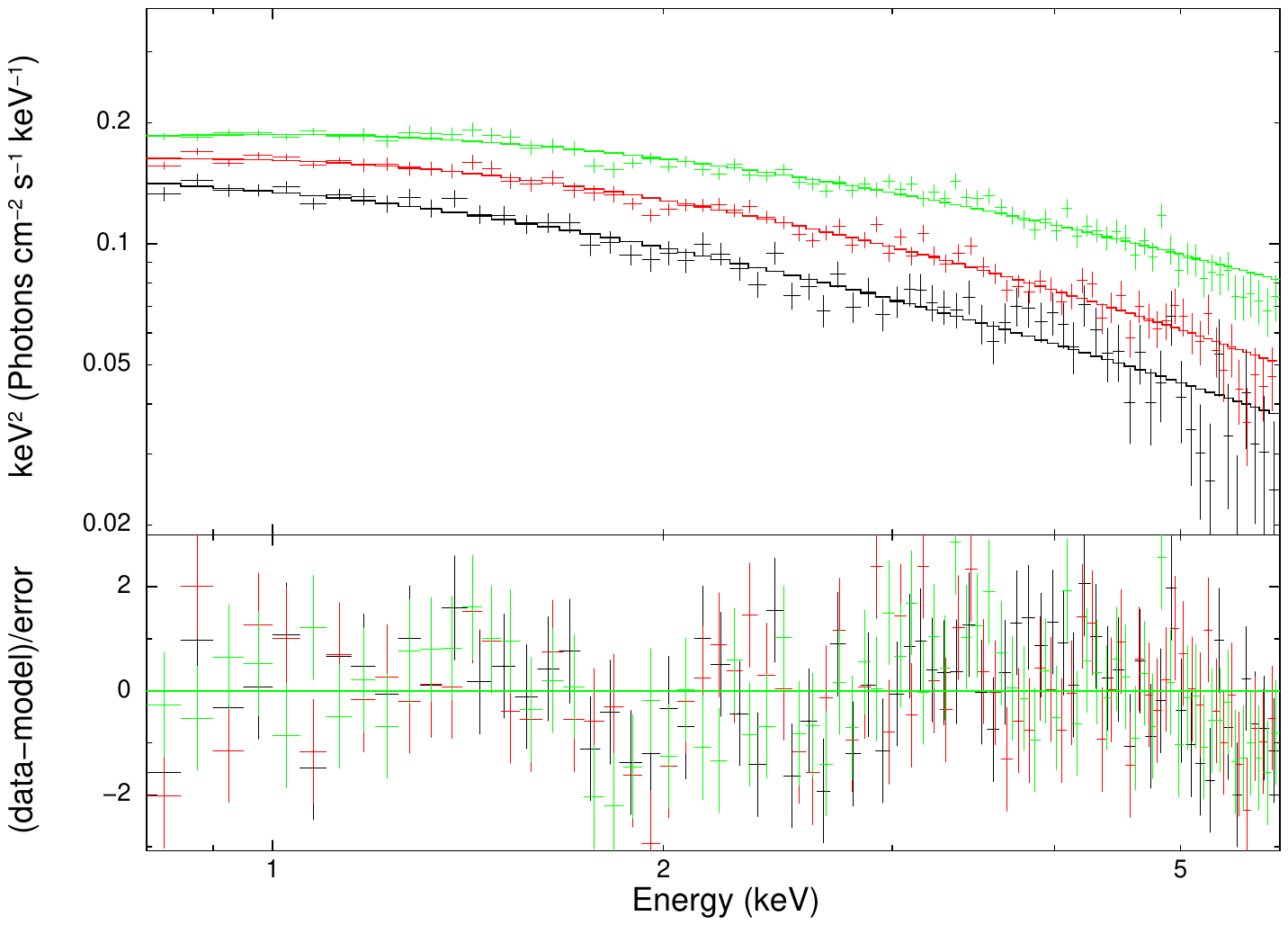}{0.55\textwidth}{(a) Log-parabola, no edge ($\chi^2_\nu=1.11$)}
  \fig{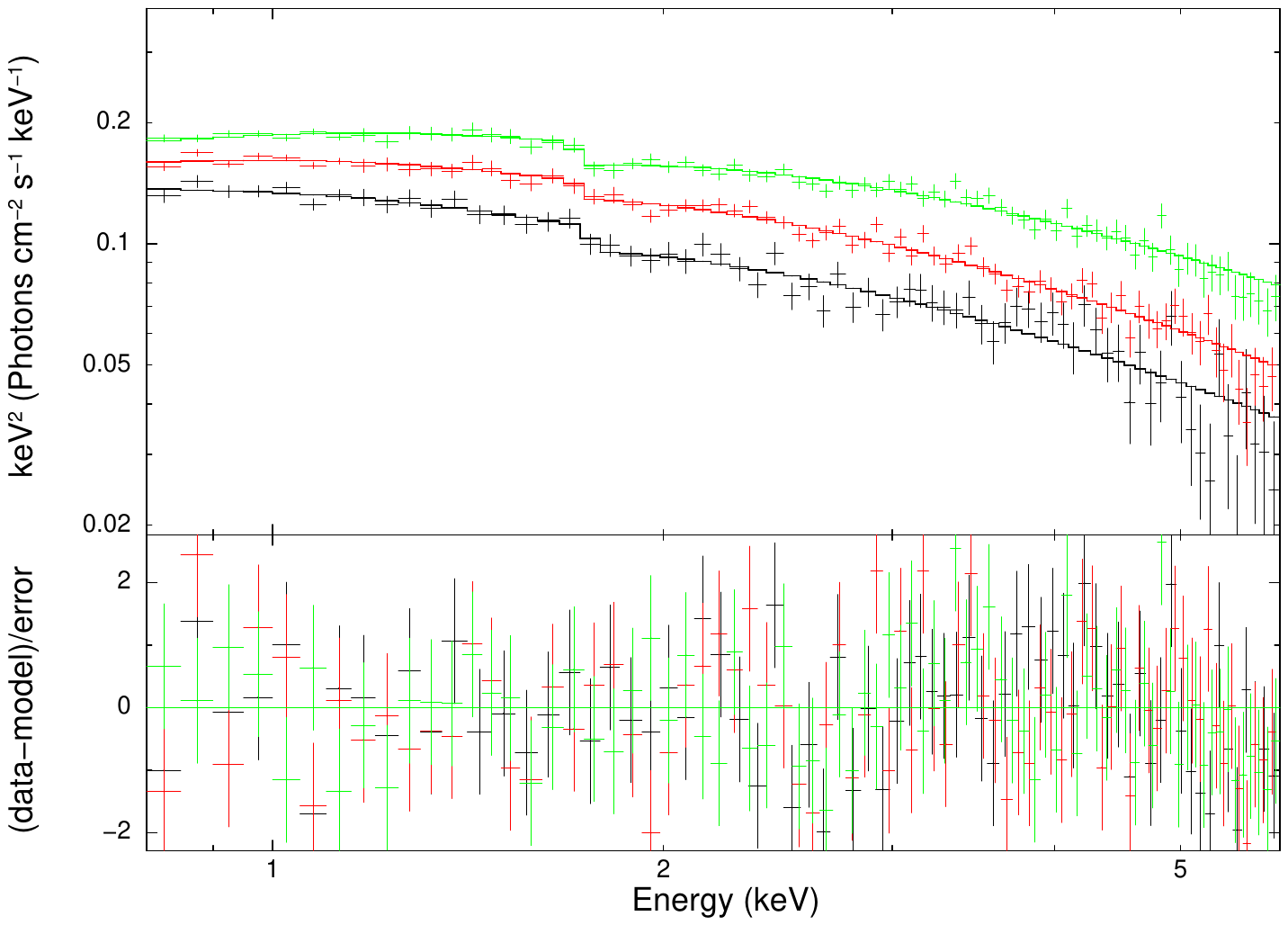}{0.55\textwidth}{(b) Log-parabola with edge ($\chi^2_\nu=0.95$)}
}
\gridline{%
  \hspace*{0.24\textwidth}%
  \fig{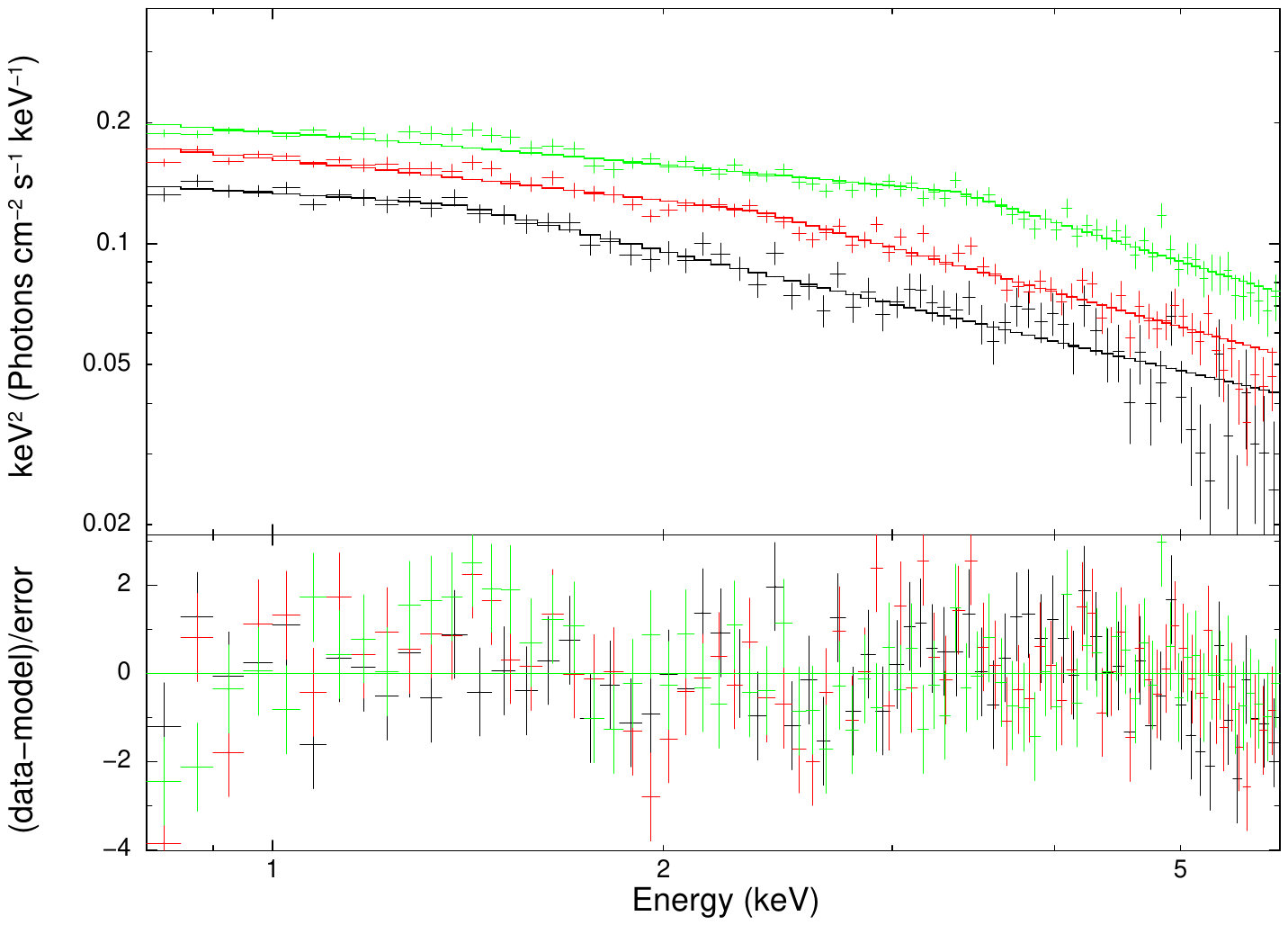}{0.55\textwidth}{(c) Broken power law, no edge ($\chi^2_\nu=1.26$)}
}
\caption{Flux-resolved XPoSat/XSPECT ($2^\circ\times2^\circ$, static background,
0.8--6~keV) spectra of Mrk~421 for the co-added low-, mid- and high-flux states,
shown as unfolded $E^2\,{\rm d}N/{\rm d}E$ spectra with fit residuals.
(a) Log-parabola without an edge: systematic residuals near 1.7~keV remain in
all three states, indicating an unmodelled absorption feature. (b) Log-parabola
with a free absorption edge: the 1.7~keV residuals are removed and the edge
energy converges to the Si~K value (1.72--1.76~keV) in every state, improving
the fit; the photon index $\alpha$ hardens from the low to the high state
(harder-when-brighter) while the curvature $\beta$ stays approximately constant.
(c) Broken power law without an edge, shown for comparison: it describes the
smoothly curved spectrum significantly less well and leaves the mid-state break
energy poorly constrained, confirming the log-parabola as the preferred
continuum model.}
\label{fig:fluxres}
\end{figure*}

To track the spectral evolution at higher signal-to-noise than the individual
days allow, we grouped the Bayesian blocks of
Section~\ref{sec:var_blocks} into three flux states by count rate: a low state
below 0.95~count~s$^{-1}$, a mid state between 0.95 and 1.20, and a high state
above 1.20. For each state a spectrum was extracted by restricting the merged
event file to the good time intervals of the constituent blocks, giving
exposures of 33.8, 89.9 and 78.2~ks. The three states were fitted jointly with
$\mathrm{edge} \times \mathrm{TBabs} \times \mathrm{logpar}$, the edge energy
and $\tau_{\rm max}$ free and independent per state; the fit is good, with
$\chi^2/{\rm dof} = 200.3/211 = 0.95$. The parameters are listed in
Table~\ref{tab:fluxres} and the spectra shown in Figure~\ref{fig:fluxres}.

\begin{figure}[htbp]
\centering
\includegraphics[width=\columnwidth]{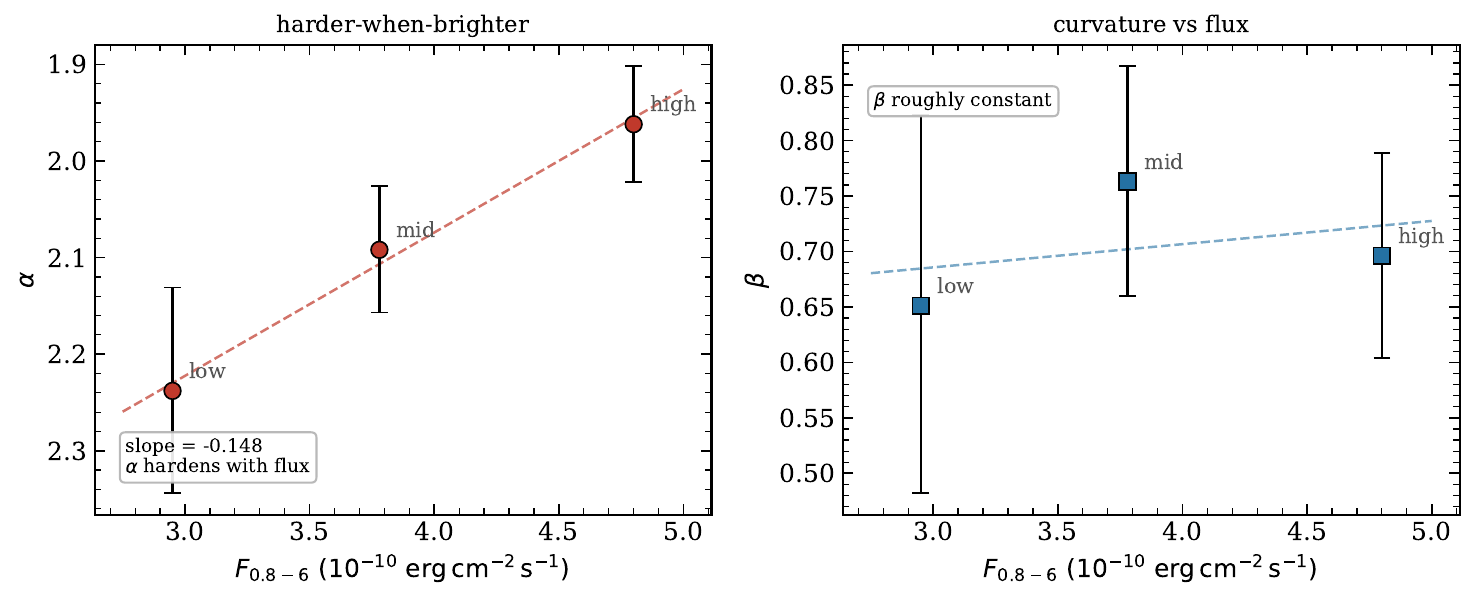}
\caption{Photon index $\alpha$ (left) and curvature $\beta$ (right) at 1~keV
against unabsorbed 0.8--6~keV flux for the three flux-resolved states. The index
hardens monotonically with flux (harder-when-brighter), the low and high states
being separated with no overlap of their confidence intervals, while the
curvature remains large and approximately constant. Error bars are 90 per cent
confidence.}
\label{fig:fluxres_ab}
\end{figure}

The edge energy converges to the Si K value in all three states (1.72–1.76 keV), with \(\tau_{\max}\approx0.10\text{–}0.14\). Omitting the edge leaves systematic residuals near 2 keV and yields \(\chi^2/\mathrm{dof}=240.2/217=1.11\), compared with \(200.3/211=0.95\) when the edge is included.
The spectrum hardens monotonically
as the source brightens: the photon index at 1~keV decreases from
$2.238^{+0.106}_{-0.107}$ in the low state through $2.092^{+0.065}_{-0.066}$ in
the mid state to $1.962^{+0.060}_{-0.060}$ in the high state (90 per cent
confidence), the low and high values separated with no overlap of their
confidence intervals. The curvature is large, $\beta \approx 0.65$--$0.76$, and
shows no systematic trend with flux, being $0.651^{+0.172}_{-0.169}$,
$0.763^{+0.104}_{-0.103}$ and $0.696^{+0.093}_{-0.092}$ from low to high. The
unabsorbed 0.8--6~keV flux rises by a factor of 1.63 across the three states,
from 2.95 to $4.80 \times 10^{-10}$~erg~cm$^{-2}$~s$^{-1}$. The decrease in photon index at approximately constant curvature indicates that the synchrotron peak shifts to higher energies as the source brightens (Figure~\ref{fig:fluxres_ab}).

We verified that the log-parabola model is statistically preferred over the broken power-law model. Fitting the same three spectral states with the \texttt{TBabs $\times$ bknpower} model (without an edge component) resulted in a significantly poorer joint fit, yielding $\chi^{2}/\mathrm{dof} = 269.6/214 = 1.26$. The discrepancy is dominated by the mid state, for which the fit statistic is $\chi^{2}/\mathrm{dof} = 113.1/73 = 1.55$. In this state, the broken power law fails to reproduce the intrinsic spectral curvature, and the break energy is driven to $E_{\rm break} \approx 2.4$~keV with a large uncertainty (Table~\ref{tab:fluxres}), as expected when a model with a sharp spectral break is applied to a smoothly curved continuum. Allowing a free absorption edge in the broken power-law model does not improve the situation. Instead, the fitted edge energy shifts away from the Si K edge, reaching approximately 5.0~keV in the low state and 1.0~keV in the mid state, indicating that the edge component is compensating for the inadequacy of the continuum model rather than accounting for the instrumental feature. We therefore adopt the log-parabola model as the baseline spectral description throughout this work.

\begin{table*}[tbp]
\centering
\caption{Flux-resolved XPoSat/XSPECT ($2^\circ\times2^\circ$, static background, 0.8--6~keV) spectral fits of Mrk~421 for the co-added low-, mid- and high-flux states. The log-parabola is fitted as \texttt{edge*TBabs*logpar} with the edge energy and $\tau_{\rm max}$ free (the edge energy converges to the Si~K value, 1.72--1.76~keV, in all states); the broken power law is fitted as \texttt{TBabs*bknpower} without an edge. The three states are fitted jointly for each model; per-state and joint $\chi^2$ are given. Without the edge the log-parabola joint fit is poorer ($\chi^2/{\rm dof}=240.2/217=1.11$ versus $200.3/211=0.95$ with the edge). $N_{\rm H}$ is frozen at $1.34\times10^{20}$~cm$^{-2}$ and the pivot at 1~keV; $F_{0.8-6}$ is the 0.8--6~keV flux in $10^{-10}$~erg~cm$^{-2}$~s$^{-1}$. Uncertainties are 90 per cent confidence.}
\label{tab:fluxres}
\begin{tabular}{ccccccc}
\hline
\multicolumn{7}{c}{\textit{Log-parabola} (\texttt{edge*TBabs*logpar}), joint $\chi^2/{\rm dof}=200.3/211=0.95$} \\
\hline
State & $F_{0.8-6}$ & $E_{\rm edge}$ (keV) & $\tau_{\rm max}$ & $\alpha$ & $\beta$ & $\chi^2/{\rm dof}$ \\
\hline
Low & $2.95$ & $1.756_{-0.095}^{+0.112}$ & $0.140_{-0.082}^{+0.083}$ & $2.238_{-0.107}^{+0.106}$ & $0.651_{-0.169}^{+0.172}$ & 62.7/66 \\
Mid & $3.78$ & $1.727_{-0.150}^{+0.128}$ & $0.096_{-0.051}^{+0.051}$ & $2.092_{-0.066}^{+0.065}$ & $0.763_{-0.103}^{+0.104}$ & 78.4/72 \\
High & $4.80$ & $1.716_{-0.072}^{+0.054}$ & $0.136_{-0.048}^{+0.049}$ & $1.962_{-0.060}^{+0.060}$ & $0.696_{-0.092}^{+0.093}$ & 59.3/73 \\
\hline
\multicolumn{7}{c}{\textit{Broken power law} (\texttt{TBabs*bknpower}, no edge), joint $\chi^2/{\rm dof}=269.6/214=1.26$} \\
\hline
State & $F_{0.8-6}$ & $\Gamma_1$ & $E_{\rm break}$ (keV) & $\Gamma_2$ & & $\chi^2/{\rm dof}$ \\
\hline
Low & $2.96$ & $2.267_{-0.189}^{+0.144}$ & $1.390_{-0.159}^{+0.228}$ & $2.751_{-0.052}^{+0.058}$ & & 72.5/67 \\
Mid & $3.77$ & $2.382_{-0.032}^{+0.078}$ & $2.377_{-0.166}^{+0.859}$ & $2.892_{-0.063}^{+0.297}$ & & 113.1/73 \\
High & $4.78$ & $2.058_{-0.089}^{+0.089}$ & $1.524_{-0.111}^{+0.174}$ & $3.015_{-0.122}^{+0.138}$ & & 84.0/74 \\
\hline
\end{tabular}
\end{table*}

\subsection{Spectral state at fixed count rate}
\label{sec:spec_rate}

The day-resolved and flux-resolved analyses group the data by count rate, which
implicitly assumes that rate is a good proxy for spectral state. The Bayesian
block decomposition lets us test that assumption directly: several blocks, well
separated in time, share the same count rate to within their errors, and we can
ask whether they also share the same spectrum. Because most individual blocks
have too few counts to constrain the log-parabola curvature, we fit each with an
absorbed power law, $\mathrm{TBabs} \times \mathrm{powerlaw}$, over 0.8--6~keV
with $N_{\rm H}$ frozen, and compare the photon index $\Gamma$ among blocks
whose rates agree within their combined $1\sigma$ uncertainty. For the short
individual blocks the static blank-sky rate exceeds the per-block source rate
and cannot be subtracted cleanly; we therefore adopt the hybrid background here.
This is the only use of the hybrid prescription in the paper, and it does not
affect the result: the test rests on the power-law index, which changes by less
than 0.07 between the two background estimates, whereas the differences reported
below reach 0.26. Blocks with fewer than 2500 net counts are excluded, since
their indices are too poorly constrained for the comparison to be meaningful;
this restricts the test to rates above about 0.9~count~s$^{-1}$.

The outcome depends on flux level. In the mid and faint regime the same rate
implies the same spectrum: four blocks at $\approx 1.14$~count~s$^{-1}$
(blocks 4, 8, 10 and 16, spanning MJD 60696--60701) share a common index to
within $2.5\sigma$ in the worst pair, and every rate-consistent pair below
1.0~count~s$^{-1}$ agrees to better than $1\sigma$. In the bright regime,
however, blocks at the same rate show significantly different indices. The
clearest case is the pair at $\approx 1.30$~count~s$^{-1}$: block~3, on an early
decline at MJD 60695.6, has $\Gamma = 2.35 \pm 0.04$, whereas block~23, on the
rising edge of the final large flare at MJD 60703.3, has
$\Gamma = 2.10 \pm 0.05$, a difference of $0.26$ at $6.4\sigma$ despite an
identical count rate eight days apart. Blocks 23 and 27 differ at $5.2\sigma$
and the pair at $\approx 1.50$ (blocks 24 and 26) at $3.5\sigma$; the pair at
$\approx 1.39$ (blocks 2 and 21) differs at $2.2\sigma$, weaker because
block~21 is faint and its index poorly constrained. The rate-consistent groups
and their indices are listed in Table~\ref{tab:spec_rate}.

\begin{table}[htbp]
\centering
\caption{Power-law index $\Gamma$ (TBabs$\times$powerlaw, 0.8--6~keV, hybrid
background) for rate-consistent groups of Bayesian blocks, testing whether count
rate uniquely labels spectral state. Groups contain blocks whose rates agree
within their combined $1\sigma$; only blocks with more than 2500 net counts are
included. Rate is in count~s$^{-1}$, errors are 90 per cent confidence.
$\Gamma_{2\times2}$ is the primary $2^\circ \times 2^\circ$ field;
$\Gamma_{3\times3}$ is the $3^\circ \times 3^\circ$ field, shown to confirm the
strongest bright-state discrepancy is field-independent. In the bright groups
(rate $\gtrsim 1.3$) the indices differ significantly, with the 3--23 and 23--27
differences reproduced in both fields; in the $\approx 1.14$ and fainter groups
they agree in the $2^\circ \times 2^\circ$ field.}
\label{tab:spec_rate}
\begin{tabular}{ccccc}
\hline
Rate group & Block & Rate & $\Gamma_{2\times2}$ & $\Gamma_{3\times3}$ \\
\hline
$\approx 1.50$ & 24 & 1.487 & $2.14 \pm 0.03$ & $2.17 \pm 0.03$ \\
               & 26 & 1.508 & $2.23 \pm 0.03$ & $2.22 \pm 0.03$ \\
\hline
$\approx 1.39$ & 2  & 1.412 & $2.35 \pm 0.04$ & $2.31 \pm 0.03$ \\
               & 21 & 1.374 & $2.24 \pm 0.08$ & $2.20 \pm 0.07$ \\
\hline
$\approx 1.30$ & 3  & 1.301 & $2.35 \pm 0.04$ & $2.28 \pm 0.03$ \\
               & 23 & 1.298 & $2.10 \pm 0.05$ & $2.15 \pm 0.05$ \\
               & 27 & 1.309 & $2.29 \pm 0.03$ & $2.28 \pm 0.03$ \\
\hline
$\approx 1.14$ & 4  & 1.122 & $2.39 \pm 0.03$ & $2.35 \pm 0.02$ \\
               & 8  & 1.151 & $2.34 \pm 0.03$ & $2.27 \pm 0.02$ \\
               & 10 & 1.141 & $2.40 \pm 0.04$ & $2.34 \pm 0.03$ \\
               & 16 & 1.151 & $2.36 \pm 0.03$ & $2.28 \pm 0.03$ \\
\hline
$\approx 0.92$ & 7  & 0.918 & $2.33 \pm 0.04$ & $2.30 \pm 0.04$ \\
               & 13 & 0.927 & $2.34 \pm 0.04$ & $2.32 \pm 0.04$ \\
\hline
\end{tabular}
\end{table}

These same-rate index differences are not specific to the primary detector set.
Repeating the per-block power-law fits on the $3^\circ \times 3^\circ$ field
reproduces the bright-regime discrepancies: blocks 3 and 23 differ by $0.12$ at
$3.3\sigma$ and blocks 23 and 27 at $4.0\sigma$. The largest discrepancy,
between blocks 3 and 23, is thus present in two independent fields with
different ancillary responses and is not a detector artefact.

The picture that emerges is that count rate is a reliable label for spectral
state when Mrk~421 is faint but not when it is bright: in the bright state,
blocks at the same rate reach genuinely different photon indices, reproduced
across both fields for the strongest pair. This is naturally understood if
bright epochs sample different phases of individual flares -- a rising, freshly
injected hard state and a declining, cooled state can pass through the same
instantaneous count rate while differing in spectral index, whereas the
fainter, more quiescent emission varies more nearly self-similarly.

\section{Discussion}
\label{sec:disc}

\subsection{Variability}

We carried out a detailed spectral study of Mrk 421 using the XSPECT and NICER observations. The analysis shows that Mrk 421 has significant variability, with
$F_{\rm var} = 19.6 \pm 0.3$ per cent over ten days and a factor of 2.3 between
the faintest and brightest resolved flux states.  The excess variance accounts for 89 per cent of the total variance in the 500~s light curve. The measured $F_{\rm var}$ is larger than the $8.6$--$15.2$ per cent reported by \citet{Markowitz2022}
from a four-day AstroSat/SXT campaign. Blazar X-ray variability
follows a red-noise process, in which most of the variability power is
carried by the slowest fluctuations, so the fractional variability measured
from a light curve depends on the range of timescales it samples; comparisons
between campaigns must therefore account for differences in baseline,
cadence, and energy band \citep{Vaughan2003}. Our ten-day baseline extends to
lower frequencies than a four-day one, capturing more of this long-timescale
power, and can therefore yield a larger $F_{\rm var}$ for that reason alone,
even if the underlying variability process is unchanged.
Separately, we noted that the data cannot constrain the size of the emitting region
through light-travel-time arguments. No variability is resolved on timescales
shorter than the 500~s sampling once the instrumental excursion at
MJD 60700.63 is removed, and fourteen of the block boundaries fall within data
gaps rather than being directly observed. This reflects the limits of our
sampling, not the source: the 23 per cent duty cycle leaves gaps of several
kiloseconds between successive exposures, so faster variability could easily
be present but unresolved.

We characterised the shape of the light curve using the rise and decay times and the asymmetry of its flare
components. Five exponential components were required to fit the XSPECT light curve, but only the last one, peaking
at MJD 60703.84, is well constrained. It has an e-folding rise time of $0.124\pm0.020$\,d (a doubling time of about
2\,h) and a decay time of $1.23\pm0.39$\,d, giving $\zeta=0.82$: a clearly asymmetric flare with a fast rise and a
slow decay. The weaker components show the same trend. The decay times of the third and fourth flares sit at the
1.5\,d bound, so these flares are at least as asymmetric as their quoted $\zeta$, while the second flare, the only
nearly symmetric one ($\zeta=0.07$), has rise and decay times that are uncertain by about 60--70 per cent. If the
variability were controlled by the light-crossing time of the emitting region, the rise and decay would be similar
and the profiles close to symmetric \citep{Chiaberge1999}. The asymmetry instead points to a strengthening and
subsequent weakening of the acceleration process \citep{2019MNRAS.484.3168S, Shah2024}. The slow decay, however,
cannot be set by radiative cooling alone. For typical jet parameters ($\delta\sim10$--20, $B\sim0.1$--0.5\,G), the
observed-frame synchrotron cooling time of the electrons emitting at 1--5\,keV is about 15\,min to 9\,h
(Section~\ref{sec:disc_spec}), shorter than $\tau_d\approx30$\,h by a factor of about 3 to 100. Only for weaker
fields, $B\lesssim0.05$\,G, does the cooling time become comparable to $\tau_d$. Because the electrons cool faster than the flux
declines, they must keep being accelerated or injected during the decay. The decay therefore traces the gradual
weakening of the acceleration rather than the cooling, and as the acceleration weakens, the synchrotron peak drifts
back to lower energies. The peak moves from about 0.66 to 1.06\,keV between the faint and bright states. This shift accounts
for only part of the brightening, about 20 per cent in count rate, and the rest must come from a rise in the peak
flux itself. The shift can, however, change the shape of the flare with energy, since the flux above a few keV
depends more strongly on the peak position than the flux near 1\,keV. In 3C~279, a similar shift of the Compton peak
gave different flare shapes in the low- and high-energy $\gamma$-ray bands \citep{2019MNRAS.484.3168S}. In Mrk~421, the harder spectrum of block~23, on the rise
of this same flare, compared with block~3 at the same count rate during a decline (Section~\ref{sec:spec_rate}),
fits this picture.

The shape of the count-rate distribution provides clues to the nature of the underlying physical processes.
A normal distribution is rejected by the AD test ($A^2=2.83$). Histogram fitting shows that a lognormal describes
the data better, and two lognormal components describe them best ($\chi^2_\nu=1.06$, against 1.69 for a single
lognormal and 1.91 for a normal). A lognormal distribution arises when the fluctuations multiply rather than add
\citep{2005MNRAS.359..345U}. This is not peculiar to blazars: lognormal fluxes are seen in the X-ray emission of
black-hole binaries and Seyfert galaxies \citep{2005MNRAS.359..345U}, in the X-ray light curve of BL~Lacertae
\citep{2009A&A...503..797G}, at very high energies in PKS~2155$-$304 \citep{2010A&A...520A..83H}, and in the
$\gamma$-ray light curves of bright \emph{Fermi} blazars \citep{2018RAA....18..141S, Shah2024}. For Mrk~421
itself, lognormality has been found in the long-term MAXI X-ray and \emph{Fermi}-LAT light curves
\citep{sinha2016long}. In accreting sources, the multiplicative behaviour is usually explained by fluctuations in the
accretion rate that arise at different radii and modulate one another as they propagate inwards
\citep{1997MNRAS.292..679L, 2005MNRAS.359..345U}. If such fluctuations are passed on to the jet, a lognormal flux
distribution would point to a connection between the disc and the jet \citep{2009A&A...503..797G,
2018RAA....18..141S}. However, this is not the only possibility. Gaussian fluctuations in the particle acceleration
timescale within the jet also give a lognormal flux distribution \citep{2018MNRAS.480L.116S}, and even an additive
process, the summed emission of many randomly oriented minijets, can produce one \citep{minijet}. In our data, the
two lognormal components have almost the same centre, near 1.1\,ct\,s$^{-1}$, but different widths: the broad
component ($\sigma=0.089$\,dex, about 23 per cent in rate) holds 84 per cent of the distribution, and the narrow
one ($\sigma=0.023$\,dex, about 5 per cent) the rest. The two components therefore do not mark two separate flux
levels. Rather, the source appears to switch between a quiet state, in which the rate barely changes, and an active
state with large flaring excursions about a similar mean. Two lognormal states have been reported before, for
example in the multiwavelength flux of PKS~1510$-$089 \citep{pankaj_ln}, and double-peaked flux distributions of
Mrk~421 in the radio and optical bands have been attributed to separate quiescent and flaring phases
\citep{sinha2016long}. In X-rays, \citet{Khatoon2020} found bimodal, probably double lognormal, long-term flux
distributions for Mrk~421 and Mrk~501 in RXTE data, and suggested that such complex distributions carry information
on the energetics of the emitting electrons. Ten days of a red-noise light curve contain only a handful of
independent excursions, however, so the split into two components should be confirmed with longer monitoring.

The power spectrum of the 0.8--7\,keV light curve follows a single power law between 0.1 and 57.6\,d$^{-1}$,
that is, on timescales from about 25\,min to 10\,d, with a slope $s=1.7$ (acceptable range 1.25--3.05). This slope
implies red-noise behaviour, in which most of the variability power is in the slowest variations; the light curve is
therefore dominated by slow changes rather than by rapid, independent fluctuations. Power spectra of this kind are
typical of blazars. In $\gamma$-rays, the slopes of bright \emph{Fermi} blazars lie mostly between 1 and 2
\citep{2010ApJ...722..520A, 2014ApJ...786..143S, bhatta2020nature}, and for BL~Lac we found a flicker-noise slope
close to 1 \citep{Shah2024}. In X-rays, \citet{Kataoka2001} found steeper spectra for Mrk~421 and other TeV
blazars, $\propto\nu^{-2}$ to $\nu^{-3}$, on timescales shorter than about a day, flattening towards longer
timescales. Our best slope, $s=1.7$, is flatter, but our frequency range extends below 1\,d$^{-1}$ into the region
where the spectrum flattens, so a single power law fitted across both regions is expected to give an intermediate
value. Our slope is therefore not directly comparable with theirs. A bending power law does not improve
our fit, so the data do not require this flattening, but they cannot exclude it either: only nine periodogram
frequencies lie below 1\,d$^{-1}$, too few to measure a change in slope given the large scatter of the periodogram.
A longer light curve would be needed to test for it. The steep slope also explains why the $F_{\rm var}$ comparison
at the start of this section must account for the baseline, since for $s\approx1.7$ the measured variance grows
strongly with the length of the light curve.

\subsection{Spectral behaviour}
\label{sec:disc_spec}
The day-resolved and flux-resolved spectra show the harder-when-brighter behaviour characteristic of high-synchrotron-peaked blazars. In the three flux-resolved states, the photon index at 1 keV decreases from approximately 2.24 in the faintest state to 1.96 in the brightest state. The anti-correlation is significant in the XSPECT-only day-wise fits (\(r=-0.83\), \(p=0.003\)) and is also present in the NICER-only fits (\(r=-0.69\), \(p=0.042\)); the joint fits have the same sense, although their correlation is not formally significant (\(r=-0.63\), \(p=0.096\); Table~\ref{tab:correlations}). The co-added flux-resolved spectra show the same monotonic evolution, with the low- and high-flux photon-index confidence intervals not overlapping. Thus, the harder-when-brighter trend is recovered in the separate XSPECT and NICER analyses and in the flux-resolved XSPECT spectra. The curvature remains large, \(\beta\approx0.7\), with no statistically significant dependence on flux. For the adopted log-parabola model, the decrease in photon index at approximately constant curvature implies that the model synchrotron peak shifts to higher energies as the source brightens. In a synchrotron interpretation, this is consistent with an increase in the characteristic electron energy, although the present data cannot distinguish changes in particle acceleration or cooling from possible changes in magnetic field or Doppler factor. The weak anti-correlation between the XSPECT-to-NICER count-rate ratio and flux has the expected sign, but is not statistically significant; it should therefore be regarded only as a consistency check, rather than independent evidence for the spectral trend.

We fitted the day-resolved XSPECT-only, NICER-only and joint XSPECT+NICER spectra, and the flux-resolved XSPECT spectra, with a log-parabola model (Section~\ref{sec:spec}). Log-parabola spectra arise naturally in models of stochastic (second-order Fermi) particle acceleration, which produce a log-parabolic electron distribution and hence a curved synchrotron spectrum. In these models the curvature parameter $\beta$ is set by the momentum-diffusion coefficient governing the acceleration process: more efficient diffusion, acting over a longer time, broadens the electron distribution and lowers the curvature \citep{tramacere2011}. For this model the synchrotron peak energy, i.e.\ the peak of the $E^2 N(E)$ spectrum, is $E_p = E_1 \times 10^{(2-\alpha)/(2\beta)}$ \citep{massaro2004log}. Evaluated at the flux-resolved fit parameters (Table~\ref{tab:fluxres}), this gives $E_p \approx 0.66$, $0.87$ and $1.06$~keV in the low, mid and high states respectively, quantifying the factor of $\sim 1.6$ peak shift. The curvature $\beta$ has been reported to decrease as the peak energy increases, both in blazars generally and in Mrk~421's own flares \citep{massaro2004log, tramacere2007signatures, tramacere2009, tramacere2011}. In the present data, however, $\beta$ is statistically indistinguishable across the three states ($0.65$--$0.76$) despite the $\sim 60$ per cent shift in $E_p$.  A shift in peak energy at constant curvature is consistent with an increase in the characteristic energy of the injected electrons, or with a change in the Doppler factor, while the acceleration process that sets $\beta$ itself remains unchanged. In either case, the spectral shape changes with flux, so the X-ray variability cannot be a change in flux normalisation alone, consistent with the pivot-energy analysis of the same source by \citet{2025JHEAp..45..438A}. Finally, we caution that the flux range spanned in this campaign (a factor of 1.63) is modest compared with the larger flares in which the curvature--peak anticorrelation was established, so the present data may simply lack the leverage to detect a change in $\beta$.

A narrow residual feature near 1.7 keV is required in the brighter log-parabola spectra of both instruments. In the flux-resolved XSPECT fits, its energy is consistently \(1.72\text{--}1.76\) keV, close to the Si K edge, and its optical depth shows no compelling dependence on flux. Its detection primarily in brighter spectra is naturally explained by their higher signal-to-noise. We therefore interpret the feature as an instrumental Si K edge associated with the silicon detector materials, rather than as absorption intrinsic to Mrk~421. The broken-power-law fits do not provide an equally satisfactory description of the spectra: when an edge is allowed to vary freely, its energy can shift away from the Si K value, indicating that it is compensating for the inadequacy of the sharp-break continuum model.

The fixed-count-rate test shows that, in the bright regime, the X-ray count rate does not uniquely determine the spectral state of Mrk~421. In other words, bright epochs with the same count rate can have significantly different spectral slopes. To compare the short Bayesian-block spectra, for which the log-parabola curvature cannot be constrained reliably, we use the power-law photon index \(\Gamma\) as an effective measure of the 0.8--6 keV spectral hardness. The clearest example occurs at a rate of approximately \(1.30\ {\rm count\ s^{-1}}\): block 3, during the early decline of the campaign, has \(\Gamma=2.35\pm0.04\), whereas block 23, on the rise of the final flare, has \(\Gamma=2.10\pm0.05\). Thus, despite having rate-consistent fluxes, the rising state is substantially harder than the declining state, with \(\Delta\Gamma=0.26\) (\(6.4\sigma\)). The same difference is recovered in the independent \(3^{\circ}\times3^{\circ}\) field of view, showing that it is not a detector or background artefact. Other rate-matched bright blocks show the same behaviour, whereas the tested mid- and low-rate groups are spectrally consistent within their uncertainties.
This result is naturally understood if the count rate traces only the band-integrated X-ray output, while the spectral slope retains information about the recent evolution of the emitting electrons. During the rise of a flare, fresh particle injection or more efficient acceleration can produce a harder spectrum; during the decay, radiative cooling and particle escape can soften the spectrum while the source passes through the same count rate. The observed differences are therefore consistent with phase-dependent spectral evolution, or spectral hysteresis, in the bright state. However, because the relevant blocks are separated by gaps and do not trace a complete individual flare, the present data do not resolve a hardness--intensity loop.

 Hardness--intensity hysteresis is a recurring feature of Mrk~421's X-ray flares: clockwise loops, in which the rising phase of a flare is harder than the declining phase at the same flux, were first reported from ASCA monitoring \citep{takahashi1996} and have since been seen repeatedly  \citep{fossati2000x, brinkmann2005, kapanadze2016}, although the amplitude of the loop vary from flare to flare. These loops are generally interpreted within kinetic models in which the finite acceleration and cooling timescales of the radiating electrons compete \citep{kirk1998}: a harder spectrum during the rise reflects fresh injection or efficient acceleration outpacing cooling, while the softer spectrum during the decline reflects the electrons cooling once injection weakens.
Blocks~3 and~23 span 35.5~ks ($\approx 9.9$~h) and 12.25~ks ($\approx 3.4$~h) from start to end, respectively (Table~\ref{tab:blocks}); their net exposures are shorter because of observational gaps. Since the source continues to evolve in these gaps, it is the elapsed span, not the net exposure, that sets how long the electron population has to evolve. For jet parameters typically adopted for Mrk~421 ($\delta \sim 10$--20, $B \sim 0.1$--0.5~G), the observed-frame synchrotron cooling time of electrons radiating at energy $E$, $t_{\rm cool,obs} \approx 9\,{\rm h}\,(B/0.1\,{\rm G})^{-3/2}(\delta/10)^{-1/2}(E/1\,{\rm keV})^{-1/2}$, ranges from about 15~min to about 9~h at 1--5~keV, shorter than or comparable to these spans over most of this range. Two consequences follow. First, the cooling time is far shorter than the more than seven days separating blocks~3 and~23, so the electrons in one block retain no memory of the other; their differing indices need not reflect any continuous evolution of the source over that interval, and each block's index may instead be set by its own recent acceleration and cooling history. Second, because the cooling time is usually shorter than or comparable to a block's span, the electron distribution can change within a single block, for example softening as freshly injected electrons cool. Each block spectrum is accumulated over its whole span, and a Bayesian block guarantees only a constant count rate, not a constant spectral shape, so the index quoted for a block may be a count-weighted average over such evolution rather than a single, fixed state.
The spectral hysteresis described above may plausibly be connected to X-ray polarization, since shock-in-jet models predict different polarization for freshly accelerated and subsequently cooled electrons at the same flux \citep{Marscher2014}. The present, X-ray-only campaign is insufficient to probe this connection; future observations with simultaneous polarimetric and spectral monitoring could check it directly.

\subsection{Cross-instrument agreement}
The agreement between XSPECT and NICER over eight quasi-simultaneous epochs, with a Pearson correlation coefficient of 0.941 and a scatter of 8.5 per cent about a proportional relation, indicates that the two instruments track the broad X-ray variability of the source consistently over the range sampled in this campaign. The relation between the two count rates established here may therefore be useful in future comparisons. Its application should, however, take account of the different XSPECT and NICER bandpasses and of possible changes in the rate ratio with spectral state, which the present eight-epoch sample is too small to test.
The strictly simultaneous coverage is limited to 254 s, so part of the observed scatter is naturally expected to arise because the two instruments can sample slightly different phases of a variable source.
 A good example of this is the one significant outlier, at MJD 60697.05, where the matched XSPECT rate is 14.6 per cent below what the fitted relation predicts (Section~\ref{sec:cross_ep3}). The Bayesian block analysis shows this is a short, real dip in the source's flux, not an instrument problem, since it appears in both XSPECT fields of view. A spectral cause is unlikely. NICER sees lower-energy photons than XSPECT, but its measured flux and spectrum at this epoch both match the campaign average. A small softening does appear in the fit, but it is much too weak to explain the size of the deficit on its own. The more likely reason is simply a mismatch in timing. NICER's pointing at this epoch lasted only 151~s, while the XSPECT block covering the dip lasted about 1.0~ks; most of NICER's short exposure therefore fell outside the dip, so its average rate barely changes because of it, while XSPECT's longer, continuous coverage catches the dip directly. The two instruments do not actually disagree about how bright the source was at any one moment -- this is just what happens when a short, real event is caught fully by one instrument's continuous coverage but only partly by the other's brief, separate pointing. This case is a useful warning for anyone using the XSPECT--NICER relation given here in future: outliers like this are expected whenever the source changes faster than the gap between the two instruments' observations, and do not mean the two instruments measure the source differently.
The weak anti-correlation between the XSPECT-to-NICER count-rate ratio and brightness has the sign expected for harder-when-brighter behaviour, but is not significant; it therefore provides only limited additional support for the spectral trend.

\section{Summary}
\label{sec:summary}

\begin{enumerate}
\item  A ten-day XSPECT campaign on Mrk 421 in 2025 January gives 202~ks of
exposure over 261 good time intervals. Detector 4 was excluded because of
localised contamination. We regenerated the responses and backgrounds for this
detector selection.

\item  The 0.8--15~keV light curve is strongly variable. The fractional
variability amplitude is $F_{\rm var} = 19.6 \pm 0.3$ per cent, and the excess
variance accounts for 89 per cent of the total.

\item  The light curve needs five exponential flare components. The one
well-constrained flare, peaking at MJD 60703.84, rises in 0.12~d and decays over
1.2~d ($\zeta=0.82$). For typical jet parameters, the decay is slower than
the synchrotron cooling of the X-ray-emitting electrons, so the electrons must keep being
accelerated during the decay, and the decay traces a gradual weakening of the acceleration.

\item  The count-rate distribution is not normal. It is best described
by two lognormal components with similar centres but different widths, which
points to multiplicative variability and to the source alternating between a
quiet and an active state.

\item  The power spectrum is a red-noise power law with slope $s=1.7$
(acceptable range 1.25--3.05) between 0.1 and 57.6~d$^{-1}$, with no evidence for
a break, as is typical of blazar variability.

\item  A Bayesian block decomposition yields 27 blocks. One is excluded as
instrumental (point 7), leaving 26 statistically distinct flux states. The
segmentation is validated against simulations of a constant source sampled on
the same time grid, which yield only one to two blocks. The resolved states
span a factor of 2.3 in count rate. Almost all of the transitions between
adjacent blocks are significant at more than $3.8\sigma$; the one exception is
the transition next to the excluded block, whose two neighbouring blocks are
consistent with the same flux level.

\item  Comparing the two fields of view identifies one interval, at MJD
60700.63, in which all eight detectors of the $2^\circ \times 2^\circ$ field of
view decline together while the $3^\circ \times 3^\circ$ field of view is
unaffected. This is instrumental, and
the corresponding block was excluded.

\item  Nine contemporaneous NICER pointings provide only 254~s of strictly
simultaneous coverage. Matching each pointing to the nearest qualifying XSPECT
interval gives eight epochs. The rates in these epochs correlate with
$r = 0.941$ ($p = 5 \times 10^{-4}$), following
$R_{\rm XSPECT} = (1.963 \pm 0.062) \times 10^{-3} R_{\rm NICER}$, with 8.5 per
cent scatter.

\item  The one significant outlier, at MJD 60697.05, has an XSPECT rate
14.6 per cent below the fitted relation. This coincides with a short, real dip
seen in both XSPECT fields of view, so it is not instrumental. A spectral
origin is disfavoured: the NICER flux at this epoch matches the campaign
average, and the small spectral softening seen in the fit is too weak to
explain the deficit alone. The more likely explanation is a timing mismatch:
NICER's brief, 151~s pointing mostly missed the dip, while XSPECT's continuous
coverage captured it directly, so the two instruments are not actually
disagreeing about the source's brightness at any single moment.

\item  Day-resolved and flux-resolved log-parabola fits show
harder-when-brighter behaviour. The photon index at 1~keV decreases from
$\approx 2.24$ in the faintest state to $\approx 1.96$ in the brightest. The
curvature stays large ($\beta \approx 0.7$) and does not depend significantly
on flux; this is consistent with the model synchrotron peak shifting to higher
energy as the source brightens, from about 0.66 to 1.06~keV. The trend appears
independently in the XSPECT-only, NICER-only, and joint fits. A broken power
law is disfavoured: it gives a significantly poorer fit and leaves the break
energy unconstrained. A narrow absorption feature near 1.7~keV, consistent
with the instrumental Si~K edge, appears in the brighter spectra of both
instruments.

\item  Comparing Bayesian blocks matched in count rate tests whether rate
uniquely labels spectral state. In the bright regime it does not: blocks at the
same rate differ in photon index by up to $0.26$ ($6.4\sigma$), a result
reproduced in both fields of view. In the mid and faint regime, blocks at
equal rate are spectrally consistent. This agrees with bright epochs sampling
different phases of individual flares at a common instantaneous rate.
\end{enumerate}

\begin{acknowledgments}
This work uses data from the XPoSat mission of the Indian Space Research
Organisation, archived at the Indian Space Science Data Centre. We thank the
XSPECT Payload Operations Centre at the U R Rao Satellite Centre for the data
processing software and calibration products. This research has made use of
data and software provided by the High Energy Astrophysics Science Archive
Research Center, a service of the Astrophysics Science Division at NASA
Goddard Space Flight Center.
\end{acknowledgments}

%

\facilities{XPoSat (XSPECT), NICER}

\software{astropy \citep{2013A&A...558A..33A, 2018AJ....156..123A,
2022ApJ...935..167A}, HEASoft \citep{2014ascl.soft08004N}, XSPEC
\citep{1996ASPC..101...17A}, NumPy \citep{2020Natur.585..357H},
Matplotlib \citep{2007CSE.....9...90H}}

\vspace{5mm}




\bibliography{sample631}{}
\bibliographystyle{aasjournal}



\end{document}